\documentclass{article}
\usepackage{amssymb,amsmath}
\usepackage{rotating}
\usepackage{graphicx,comment}
\usepackage{color}
\usepackage{caption}
\usepackage{subcaption}
\usepackage[margin=1.5in]{geometry}
\pdfoutput=1

\newtheorem{theorem}{Theorem}

\def\grad{{\bf \nabla}}
\def\div{{\bf \nabla\cdot}}

\usepackage{authblk}

\title{Time-dependent injection strategies and interfacial stability in multi-layer Hele-Shaw and porous media flows}

\author[1]{Craig Gin}
\author[2]{Prabir Daripa}
\affil[1]{Department of Applied Mathematics, University of Washington, 
Seattle, WA 98195}
\affil[2]{Department of Mathematics, Texas A\&M University, 
College Station, TX 77843}

\date{}

\begin{document}

\maketitle

\begin{abstract}
We study the stability of multi-layer radial flows in porous media within the Hele-Shaw model. We perform a linear stability analysis for radial flows consisting of an arbitrary number of fluid layers with interfaces separating fluids of constant viscosity and with positive viscosity jump at each interface in the direction of flow. Several different time-dependent injection strategies are analyzed including the maximal injection rate that maintains a stable flow. We find numerically that flows with more fluid layers can be stable with faster time-dependent injection rates than comparable flows with fewer fluid layers. In particular, the injection rate for a stable flow increases at a rate that is proportional to the number of interfaces to the two-thirds power for large times. Additionally, we show that in any multi-layer radial Hele-Shaw flow, if all of the interfaces are circular except for one perturbed interface then there exists a time-dependent injection rate such that the circular interfaces remain circular as they propagate and the disturbance on the perturbed interface decays. 

The motion of the interfaces within linear theory is also investigated numerically for the case of constant injection rates. It is found that: (i) A disturbance of one interface can be transferred to the other interface(s); (ii) The disturbances on the interfaces can develop either in phase or out of phase from any arbitrary initial disturbance; and (iii) The dynamics of the flow can change dramatically with the addition of more interfaces.

\end{abstract}

\section{Introduction}
There are many applications in which one fluid displaces another fluid in a porous medium including oil recovery, hydrology, filtration, and fixed bed regeneration in chemical processing. In the case that the displacing fluid is less viscous than the displaced fluid, the interface between the fluids is unstable and viscous fingering ensues. 
In some of the aforementioned applications, several fluids are used in succession to displace the resident fluid in the porous medium. An example of this is chemical enhanced oil recovery (EOR) in which a sequence of complex fluids with favorable properties are injected into an oil reservoir. If the properties of these fluids and the injection procedure are chosen 
properly then there can be significant improvement in oil recovery over water flooding methods. It is well-known that viscous fingering limits the displacement efficiency in such processes so controlling this instability is very important for industrial applications. In what follows, we consider the case in which the injected fluids are all immiscible. In chemical EOR, it is commonplace to use a variety of fluids in succession to flood the reservoir. In some flooding schemes, it is possible that an aqueous phase based liquid is displacing a different aqueous phase liquid. In such a case, a thin layer of spacer fluid of some non-aqueous phase liquid (NAPL) between such miscible phases can be used to ensure immiscibility \cite{daripa-gin:dispersion}. Alternatively, the injected fluid layers themselves can alternate between aqueous phases and NAPL's. 

Viscous fingering in porous media flows \cite{Homsy:1987} is often studied using the Hele-Shaw model in which there is a sharp interface between immiscible fluids. A linear stability analysis of viscous fingering within the Hele-Shaw model was first performed by Saffman and Taylor \cite{Saffman/Taylor:1958}. They studied the case in which the fluid moves linearly and orthogonal to a planar interface. We refer to this flow configuration as rectilinear flow. 
An appropriate model for flow near an injection or production well is radial flow. 
The stability of radial Hele-Shaw flows was first studied by Bataille \cite{Bataille:1968} and Wilson \cite{Wilson:1975} and was later developed further by Paterson \cite{Paterson:1981}.
Recently there have been many studies of the stability of Hele-Shaw flows with more complex physics in both the rectilinear and radial geometries. These include tapered Hele-Shaw cells \cite{Al-Housseiny/Stone:2013}, inertial effects \cite{Dias/Miranda:2011}, and flows of chemically reactive \cite{He/Lowengrub/Belmonte:2012} or non-Newtonian fluids 
\cite{Coussot:1999,Fontana/Dias/Miranda:2014}, but almost all involve two layers of fluids separated by one interface initially. 

There have been relatively few studies on flows with more than two fluid layers. Cardoso and Woods 
\cite{Cardoso/Woods:1995} studied three-layer flows in both the rectilinear and radial flow geometries. For radial flows, they considered flows in which the inner interface is stable and used the linear theory to predict the number of drops formed when the interfaces meet. Beeson-Jones and Woods \cite{Beeson-Jones/Woods:2015} studied three-layer radial flow
and found the optimal value of the viscosity of the intermediate fluid in order to inject fluid at the fastest rate possible while maintaining a stable flow. Daripa and various collaborators have studied various aspects of multi-layer flows in both the rectilinear and radial geometries. 
In a paper by Daripa \cite{daripa08:studies}, three-layer rectilinear Hele-Shaw flows are studied and a formula is given for a critical value of the viscosity of the middle layer fluid that minimizes the bandwidth of unstable waves. Daripa also formulated the stability problem for rectilinear Hele-Shaw flows with an arbitrary number of fluid layers \cite{daripa08:multi-layer}. In that paper, the Rayleigh quotient and some inequalities are used to derive upper bounds on the growth rate of instabilities.
Similar results for multi-layer radial flow are given by Gin and Daripa \cite{gin-daripa:hs-rad}.

In controlling the instability of Hele-Shaw and porous media flows, one of the simplest parameters to control is the rate at which fluid is injected. 
There have been several studies on controlling the instability by using a time-dependent injection rate. The most dangerous wave number (and hence the number of fingers) can be made constant in time by considering an injection rate that scales with time like $t^{-1/3}$. This type of injection rate was studied numerically by Li et al. \cite{Li/Lowengrub/etc:2009} and the limiting shape of the interface is found to be independent of the initial conditions. Zheng et al. 
\cite{Stone/etc:2015} studied a class of time-dependent control strategies of which the $t^{-1/3}$ injection rate is a special case.  Dias et al. \cite{Dias/etc:2012} studied the optimal injection policy when a given amount of fluid needs to be pumped in a given amount of time and found that a linearly increasing injection rate is optimal. Beeson-Jones and Woods \cite{Beeson-Jones/Woods:2015} found the maximum possible time-dependent injection rate for a stable flow analytically for two-layer flow and numerically for three-layer flow. In the present work, we extend some of these results on time-dependent injection to flows with three or more fluids. We also study the linearized motion of interfaces under constant as well as time-dependent injection rates.

The paper is laid out as follows. In \S\ref{sec:Preliminaries}, the stability problem is formulated for multi-layer Hele-Shaw flows with an arbitrary number of fluid layers. Several different time-dependent injection strategies that can be used to control the instability 
are described and analyzed in \S\ref{sec:TimeDepQ}. Numerical results are given in \S\ref{sec:Numerical} and then concluding remarks are given in \S\ref{sec:Conclusion}.

\section{Preliminaries}\label{sec:Preliminaries}
\begin{figure}
\centering
\includegraphics[width=2.5 in,height=2 in]{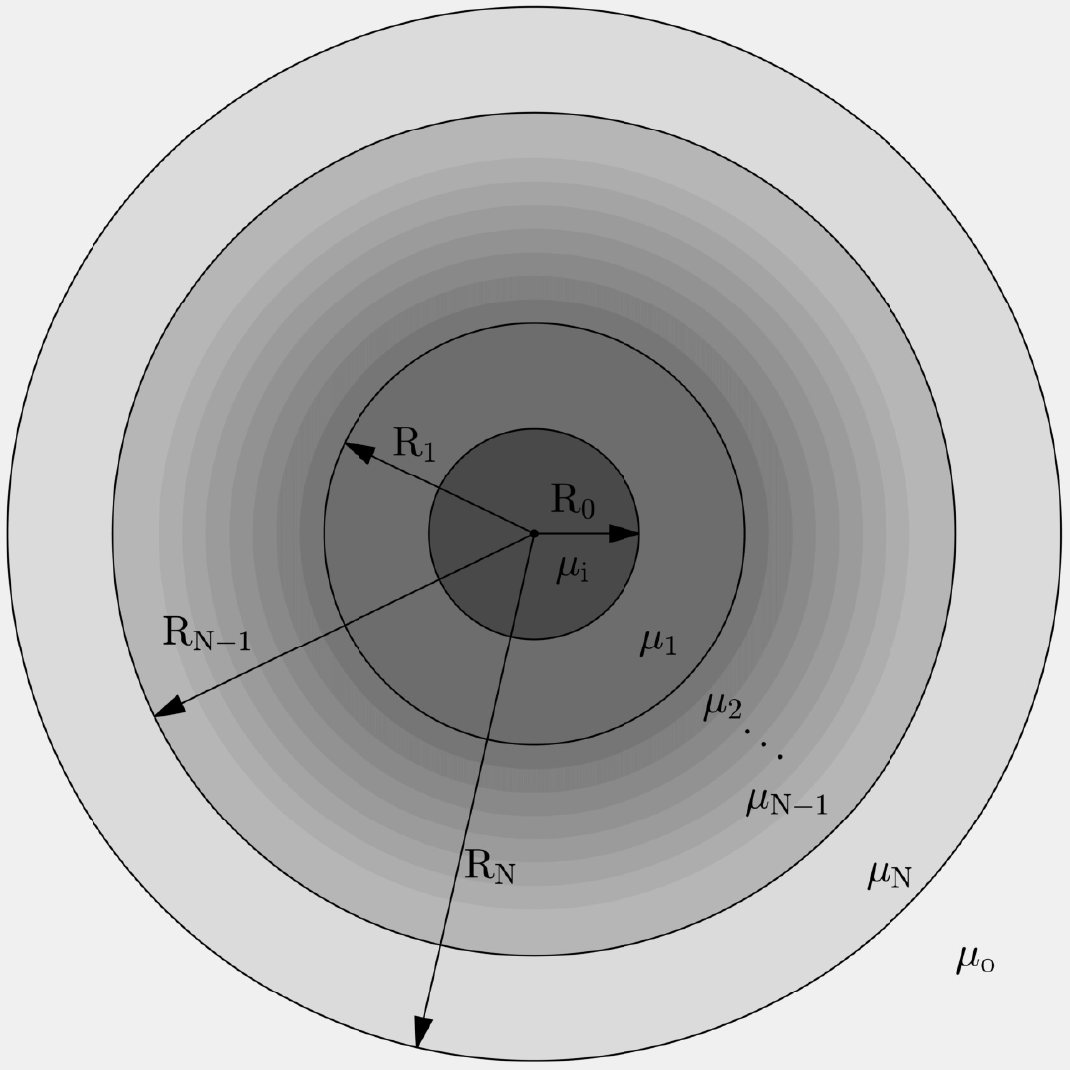}
\caption{The basic solution for (N+2)-layer flow}
\label{Constant:3Layer_Initial}
\end{figure}

Consider a radial Hele-Shaw flow consisting of $N+2$ regions of incompressible, immiscible fluid. By averaging across the gap, we may consider two-dimensional flow in polar coordinates, $\Omega := (r,\theta) = \mathbb{R}^{2}$. The least viscous fluid with viscosity $\mu_i$ is injected into the center of the cell at injection rate $Q$. The most viscous fluid, with viscosity $\mu_o$, is the outermost fluid. There are $N$ internal layers of fluid with viscosity $\mu_j$ for $j = 1, \hdots, N$ where $\mu_i < \mu_j < \mu_o$ for all $j$.
The fluid flow is governed by the following equations
\begin{equation}\label{Constant:eq:main}
 \div{\bf u} = 0,\qquad \grad\;p=-\frac{\mu}{\kappa}\;{\bf u},  \qquad \text{for } r \neq 0,
\end{equation}
where $\kappa = b^2/12$, and $b$ is the width of the gap of the Hele-Shaw cell. The first equation $\eqref{Constant:eq:main}_1$ is the continuity equation for incompressible flow, and the second equation $\eqref{Constant:eq:main}_2$ is Darcy's law \cite{Darcy:1856}. Initially, the fluids are separated by circular interfaces with radii $R = R_j(0)$, $j = 0,...,N$, where $R_j(t)$ are the positions of the interfaces at time $t$, and $T_j$ are the corresponding interfacial tensions. This set-up is 
shown in Figure \ref{Constant:3Layer_Initial}.

The equations admit a simple basic solution in which all of the fluid moves outward radially with velocity $\mathbf{u} := \left(u_r,u_{\theta}\right) = \left(Q/(2 \pi r),0\right)$. The 
interfaces remain circular and move outward with velocity $Q/(2 \pi R_j(t))$. The pressure, $p = p(r)$, may be obtained by integrating equation $\eqref{Constant:eq:main}_2$.

We scale the variables using the characteristic length $R_N(0)$, the characteristic interfacial tension $T_N$, and the characteristic viscosity $\mu_o$. The characteristic injection rate $Q_{ref}$ is the injection rate at which a single interface at $R_N(0)$ with interfacial tension $T_N$ and for which a fluid with viscosity $\mu_o$ is displaced by an inviscid fluid becomes unstable to a disturbance with wave number 2 (see \cite{Beeson-Jones/Woods:2015}). Therefore,
\begin{align}
 &Q_{ref} = \frac{12 \pi \kappa T_N}{R_N(0) \mu_o}, \\ 
 &Q^* = \frac{Q}{Q_{ref}}, \\ 
 &r^* = \frac{r}{R_N(0)}, \\
 &\mu^* = \frac{\mu}{\mu_o}, \\ 
 &t^* = \frac{12 \kappa T_N}{R_N^3(0) \mu_o} t, \\
 &\mathbf{u}^* = \frac{R_N^2(0) \mu_o}{12 \kappa T_N} \mathbf{u}, \\
 &p^* = \frac{R_N(0)}{12 T_N} p, \\
 &T^* = \frac{T}{T_N}.
\end{align}
In these dimensionless variables, equation \eqref{Constant:eq:main} becomes
\begin{equation}\label{Constant:eq:main_dimless}
 \nabla^* \cdot {\bf u^*} = 0,\qquad \grad^*p^*=-\mu^*{\bf u^*},  \qquad \text{for } r^* \neq 0,
\end{equation}
and the velocity of the basic solution is ${\bf u^*} = (Q^*/(2 r^*),0)$. With a slight abuse of notation, we drop the stars below for convenience. Minimal details on the linear stability of this basic solution is given below in order to derive the maximum time-dependent injection rate for stable multi-layer flows and to develop the dynamical system governing the linearized motion of many interfaces which is subsequently studied analytically and numerically.

We perturb the basic solution $\left(u_r, u_{\theta}, p \right)$ by $\left(\tilde{u}_r,\tilde{u}_{\theta},\tilde{p}\right)$.
Since equations \eqref{Constant:eq:main_dimless} are linear, the disturbances satisfy the same equations. Therefore,
\begin{equation}\label{Constant:lindist}
\frac{\partial \tilde{u}_r}{\partial r} + \frac{\tilde{u}_r}{r} + \frac{1}{r} \frac{\partial \tilde{u}_{\theta}}{\partial \theta} = 0,\quad
\frac{\partial \tilde{p}}{\partial r} = -\mu \tilde{u}_r,\quad
\frac{1}{r} \frac{\partial \tilde{p}}{\partial \theta} = - \mu \tilde{u}_{\theta}.
\end{equation}
We use separation of variables and assume that the disturbances are of the form
\begin{equation}\label{Constant:SepVariables}
(\tilde{u}_r,\tilde{u}_{\theta},\tilde{p}) = (f(r), \tau(r), \psi(r))g(t) e^{in\theta}.
\end{equation}
Using \eqref{Constant:SepVariables} in equation $\eqref{Constant:lindist}$ yields the following ordinary differential equation for $f(r)$:
\begin{equation}\label{RadEigenvalue2}
 \left(r^3f'(r)\right)' - \left(n^2 - 1\right) r f(r) = 0.
\end{equation}

The above equation is exact meaning there has been no linearization of any sort up until now. Next we derive the boundary conditions for this equation from linearization of the interfacial dynamic boundary conditions. Let the disturbance of the interface located at $R_j$ be given by $A_n^j(t) e^{in \theta}$. The linearized kinematic interface conditions are given by
\begin{equation}\label{multi:RadkinBC_General}
 \frac{d A_n^j(t)}{dt} = f(R_j)g(t) - A_n^j(t) \frac{Q}{2 R_j^2}.
\end{equation}
The linearized dynamic interface condition (see equation (10) in \cite{gin-daripa:hs-rad}) at the innermost interface located at $R = R_0$ is 
\begin{equation}\label{multi:RaddynBCinner}
\begin{split}
   &\Big\{ f(R_0) (\mu_i - \mu_1) + R_0 \big[ \mu_i (f^-)'(R_0) - \mu_1 (f^+)'(R_0) \big] \Big\} g(t) \\
 = &\left\{\frac{Q n^2}{2 R_0^2} (\mu_1 - \mu_i) - \frac{T_0}{12} \frac{n^4 - n^2}{R_0^3} \right\} A_n^0(t).
\end{split}
\end{equation}
The linearized dynamic condition at the outermost interface located at $R = R_N$ is 
\begin{equation}\label{multi:RaddynBCouter}
\begin{split}
   &\Big\{ f(R_N) (\mu_N - 1) + R_N \big[ \mu_N (f^-)'(R_N) - (f^+)'(R_N) \big] \Big\} g(t) \\
 = &\left\{\frac{Q n^2}{2 R_N^2} (1 - \mu_N) - \frac{1}{12} \frac{n^4 - n^2}{R_N^3} \right\} A_n^N(t).
\end{split}
\end{equation}
For any intermediate interface at $R = R_j$, the dynamic condition is
\begin{equation}\label{multi:RaddynBCintermediate}
\begin{split}
   &\Big\{ f(R_j) (\mu_j - \mu_{j+1}) + R_j \big[ \mu_j (f^-)'(R_j) - \mu_{j+1} (f^+)'(R_j) \big] \Big\} g(t) \\
 = &\left\{\frac{Q n^2}{2 R_j^2} (\mu_{j+1} - \mu_j) - \frac{T_j}{12} \frac{n^4 - n^2}{R_j^3} \right\} A_n^j(t).
\end{split}
\end{equation}
Using the fact that $f(r)$ is a solution to the differential equation \eqref{RadEigenvalue2}, the dynamic interface condition \eqref{multi:RaddynBCinner} at the innermost interface becomes
\begin{equation}\label{multi:InnerDyn}
   \left\{ \mu_i - \mu_1 \frac{\left(\frac{R_0}{R_1}\right)^{2n} + 1}{\left(\frac{R_0}{R_1}\right)^{2n} - 1} \right\} f(R_0) g(t) +2 \mu_1 \frac{\left(\frac{R_0}{R_1}\right)^{n-1}}{\left(\frac{R_0}{R_1}\right)^{2n}-1} f(R_1) g(t) = F_0 A_n^0(t),
\end{equation}
where
\begin{equation}\label{multi:F0}
 F_0 = \frac{Q n}{2 R_0^2} (\mu_1 - \mu_i) - \frac{T_0}{12} \frac{n^3 - n}{R_0^3}.
\end{equation}
Similarly, the dynamic interface condition \eqref{multi:RaddynBCouter} for the outermost interface reduces to
\begin{equation}\label{multi:OuterDyn}
   \left\{ 1 - \mu_N \frac{\left(\frac{R_{N-1}}{R_N}\right)^{2n} + 1}{\left(\frac{R_{N-1}}{R_N}\right)^{2n} - 1} \right\} f(R_N) g(t) +2 \mu_N \frac{\left(\frac{R_{N-1}}{R_N}\right)^{n+1}}{\left(\frac{R_{N-1}}{R_N}\right)^{2n}-1} f(R_{N-1}) g(t) = F_N A_n^N(t),
\end{equation}
where
\begin{equation}\label{multi:F1}
 F_N = \frac{Q n}{2 R_N^2} (1 - \mu_N) - \frac{1}{12} \frac{n^3 - n}{R_N^3},
\end{equation}
and the dynamic interface conditions at the intermediate interfaces are
\begin{equation}\label{multi:IntermedDyn}
\begin{split}
  &\left\{ -\mu_j \frac{\left(\frac{R_{j-1}}{R_j}\right)^{2n} + 1}{\left(\frac{R_{j-1}}{R_j}\right)^{2n}-1} -\mu_{j+1} \frac{\left(\frac{R_{j}}{R_{j+1}}\right)^{2n} + 1}{\left(\frac{R_{j}}{R_{j+1}}\right)^{2n}-1}\right\} f(R_j)g(t)  \\
  &+2\mu_j \frac{\left(\frac{R_{j-1}}{R_j}\right)^{n+1}}{\left(\frac{R_{j-1}}{R_{j}}\right)^{2n}-1}f(R_{j-1})g(t) + 2 \mu_{j+1}    \frac{\left(\frac{R_{j}}{R_{j+1}}\right)^{n-1}}{\left(\frac{R_{j}}{R_{j+1}}\right)^{2n}-1} f(R_{j+1})g(t) \\
   = & F_j A_n^j(t),
\end{split}
\end{equation}
where
\begin{equation}\label{multi:Fj}
 F_j = \frac{Q n}{2 R_j^2} (\mu_{j+1} - \mu_j) - \frac{T_j}{12} \frac{n^3 - n}{R_j^3}, \qquad j = 1,\dots,N-1.
\end{equation}
Equations \eqref{multi:InnerDyn}, \eqref{multi:OuterDyn}, and \eqref{multi:IntermedDyn} can be written as a system of equations of the form
\begin{equation*}
 \mathbf{\widetilde{M}}_N(t)  \left( \begin{array}{c} f(R_0)g(t) \\ \vdots \\ f(R_N) g(t) \end{array} \right)  =  \left( \begin{array}{c} F_0 A_n^0(t) \\ \vdots \\ F_N A_n^N(t) \end{array} \right),
\end{equation*}
where $\mathbf{\widetilde{M}}_N(t)$ is the $(N+1) \times (N+1)$ tridiagonal matrix with entries indexed by $i,j = 0,...N$ given by
\begin{equation}\label{M_N_tilde}
\begin{split}
 &\left(\mathbf{\widetilde{M}}_N(t)\right)_{00} =  \mu_i - \mu_1 \frac{\left(\frac{R_0}{R_1}\right)^{2n} + 1}{\left(\frac{R_0}{R_1}\right)^{2n} - 1}, \qquad \left(\mathbf{\widetilde{M}}_N(t)\right)_{01} = 2 \mu_1 \frac{\left(\frac{R_0}{R_1}\right)^{n-1}}{\left(\frac{R_0}{R_1}\right)^{2n}-1} \\
 &\left(\mathbf{\widetilde{M}}_N(t)\right)_{j,j-1} = 2\mu_j \frac{\left(\frac{R_{j-1}}{R_j}\right)^{n+1}}{\left(\frac{R_{j-1}}{R_{j}}\right)^{2n}-1}, \\
 &\left(\mathbf{\widetilde{M}}_N(t)\right)_{j,j} =  -\mu_j \frac{\left(\frac{R_{j-1}}{R_j}\right)^{2n} + 1}{\left(\frac{R_{j-1}}{R_j}\right)^{2n}-1} -\mu_{j+1} \frac{\left(\frac{R_{j}}{R_{j+1}}\right)^{2n} + 1}{\left(\frac{R_{j}}{R_{j+1}}\right)^{2n}-1}, \\
 &\left(\mathbf{\widetilde{M}}_N(t)\right)_{j,j+1} = 2 \mu_{j+1}    \frac{\left(\frac{R_{j}}{R_{j+1}}\right)^{n-1}}{\left(\frac{R_{j}}{R_{j+1}}\right)^{2n}-1}, \\
 &\left(\mathbf{\widetilde{M}}_N(t)\right)_{N,N-1} = 2 \mu_N \frac{\left(\frac{R_{N-1}}{R_N}\right)^{n+1}}{\left(\frac{R_{N-1}}{R_N}\right)^{2n}-1}, \qquad \left(\mathbf{\widetilde{M}}_N(t)\right)_{N,N} = 1 - \mu_N \frac{\left(\frac{R_{N-1}}{R_N}\right)^{2n} + 1}{\left(\frac{R_{N-1}}{R_N}\right)^{2n} - 1} . 
 \end{split}
\end{equation}
Combining this system of equations with the linearized kinematic interface conditions yields the dynamical system governing the evolution of interfacial disturbances for multi-layer flows.
\begin{equation}\label{MultiLayerMatEqn}
 \frac{d}{dt} \left( \begin{array}{c} A_n^0(t) \\ \vdots \\ A_n^N(t) \end{array} \right) = \mathbf{M}_N(t) \left( \begin{array}{c} A_n^0(t) \\ \vdots \\ A_n^N(t) \end{array} \right),
\end{equation}
where
\begin{equation}\label{M}
 \textbf{M}_N(t) =  \mathbf{\widetilde{M}}_N^{-1}(t) \left( \begin{array}{ccc} F_0 &  \dots & 0 \\ \vdots & \ddots & \vdots \\ 0 & \dots & F_N \end{array} \right)  - \frac{Q}{2} \left( \begin{array}{ccc} \frac{1}{R_0^2} &  \dots & 0 \\ \vdots & \ddots & \vdots \\ 0 & \dots & \frac{1}{R_N^2} \end{array} \right) 
\end{equation}
For the particular case of $N = 1$ (i.e. three-layer flow), the matrix $\mathbf{\widetilde{M}}_1$ is a $2 \times 2$ matrix and can be easily inverted and plugged into equation \eqref{M} to obtain the matrix $\mathbf{M}_1(t)$ with entries indexed by $i,j = 0,1$ given by
\begin{equation}\label{M_3Layer}
\begin{split}
 \Big(\mathbf{M}_1(t)\Big)_{00} &=  \frac{\left\{(1 + \mu_1) - (1 - \mu_1) \left(\frac{R_0}{R_1}\right)^{2n} \right\}  F_0 }{(\mu_1 - \mu_i) (1 - \mu_1) \left(\frac{R_0}{R_1}\right)^{2n} + (\mu_1 + \mu_i) (1 + \mu_1)} - \frac{Q}{2 R_0^2}, \\
 \Big(\mathbf{M}_1(t)\Big)_{01} &=  \frac{2 \mu_1 \left(\frac{R_0}{R_1}\right)^{n-1} F_1}{(\mu_1 - \mu_i) (1 - \mu_1) \left(\frac{R_0}{R_1}\right)^{2n} + (\mu_1 + \mu_i) (1 + \mu_1)}, \\
 \Big(\mathbf{M}_1(t)\Big)_{10} &= \frac{2 \mu_1 \left(\frac{R_0}{R_1}\right)^{n+1} F_0}{(\mu_1 - \mu_i) (1 - \mu_1) \left(\frac{R_0}{R_1}\right)^{2n} + (\mu_1 + \mu_i) (1 + \mu_1)}, \\
 \Big(\mathbf{M}_1(t)\Big)_{11} &=  \frac{\left\{(\mu_1 + \mu_i) + (\mu_1 - \mu_i) \left(\frac{R_0}{R_1}\right)^{2n} \right\}  F_1 }{(\mu_1 - \mu_i) (1 - \mu_1) \left(\frac{R_0}{R_1}\right)^{2n} + (\mu_1 + \mu_i) (1 + \mu_1)} - \frac{Q}{2 R_1^2}.
\end{split}
\end{equation}
Note that these entries are time-dependent because the radii $R_0$ and $R_1$ are time-dependent as well as possibly the injection rate $Q$. Hence the matrix $\mathbf{M}_1(t)$ is time-dependent. Accounting for differences in notation and scaling, this matrix agrees with the one derived in \cite{Beeson-Jones/Woods:2015}.

\section{Time-dependent injection rate}\label{sec:TimeDepQ}
Traditionally, the injection rate $Q$ for a radial Hele-Shaw flow is taken to be constant. However, in the formulation above, the injection rate can be a function of time. There have been many studies that have explored the implications of using a strategically chosen time-dependent injection rate (see for example \cite{Beeson-Jones/Woods:2015}, \cite{Dias/etc:2012}, \cite{Li/Lowengrub/etc:2009}, and \cite{Stone/etc:2015}). Almost all of these studies are for two-layer flows with the exception of \cite{Beeson-Jones/Woods:2015} which considers three-layer flows. {In this section, we extend some of these results to flows with an arbitrary number of layers and also present some new results for time-dependent injection rates for three-layer flows. 

\subsection{Maximum Injection Rate for a Stable Flow}
First, we explore the time-dependent maximum injection rate for which a particular flow is stable. This is important for many applications including oil recovery in which it is beneficial to stabilize the flow, but it is also cost effective to inject as fast as possible. 

\subsubsection{Two-layer flow}\label{sec:Q(t)_2-layer}
In order to set the stage for time-dependent injection rates for flows with many fluid layers, we review a result that has already been established for two-layer flow. As derived in \cite{Beeson-Jones/Woods:2015}, the maximum dimensionless injection rate for which the disturbance with wave number $n$ is stable is 
\begin{equation}\label{MaxStableQ_2Layer}
 Q_M(n)  = \frac{1}{6R} \frac{n\left(n^2-1\right)}{n (1 - \mu_i) - (1+\mu_i)}.
\end{equation}
The maximum injection rate for which the flow is stable, which we denote by $Q_M$, is found by taking the minimum of equation $\eqref{MaxStableQ_2Layer}$ over all values of $n$. It is shown in \cite{Beeson-Jones/Woods:2015} that $Q_M \propto t^{-1/3}$ for $t >> 1$.

\subsubsection{Three-layer Flow}\label{sec:MaxQ_3Layer}
We now wish to obtain a result analogous to equation \eqref{MaxStableQ_2Layer} but for multi-layer flows. An exact expression like equation \eqref{MaxStableQ_2Layer} is not feasible in this case, so instead we look for bounds that ensure stability of the flow.

We start by considering three-layer flow. By Gershgorin's Circle Theorem, both of the eigenvalues of $\mathbf{M}_1$ will have a negative real part if the terms on the diagonal are negative and greater in absolute value than the off-diagonal terms in the same row. This condition is satisfied if the following two inequalities hold
\small
\begin{equation}\label{Gershgorin}
\begin{split}
 \frac{Q}{2 R_0^2} \geq&  \frac{\left\{(1 + \mu_1) - (1 - \mu_1) \left(\frac{R_0}{R_1}\right)^{2n} \right\}  F_0 +2 \mu_1 \left(\frac{R_0}{R_1}\right)^{n-1} F_1}{(\mu_1 - \mu_i) (1 - \mu_1) \left(\frac{R_0}{R_1}\right)^{2n} + (\mu_1 + \mu_i) (1 + \mu_1)}, \\ 
 \frac{Q}{2 R_1^2} \geq&  \frac{\left\{(\mu_1 + \mu_i) + (\mu_1 - \mu_i) \left(\frac{R_0}{R_1}\right)^{2n} \right\}  F_1 + 2 \mu_1 \left(\frac{R_0}{R_1}\right)^{n+1} F_0}{(\mu_1 - \mu_i) (1 - \mu_1) \left(\frac{R_0}{R_1}\right)^{2n} + (\mu_1 + \mu_i) (1 + \mu_1)}. 
\end{split}
\end{equation}
\normalsize
Recall from equations \eqref{multi:F0} and \eqref{multi:F1} that both $F_0$ and $F_1$ depend on $Q$. By using \eqref{multi:F0} and \eqref{multi:F1} in equation \eqref{Gershgorin} and solving for $Q$, the following condition on the injection rate is obtained which is sufficient to ensure that the eigenvalues of $\mathbf{M}_1$ have negative real parts.
\begin{equation}\label{3-LayerUB}
  Q \leq \min\{G_0(n),G_1(n)\},
\end{equation}
where 
\begin{align}
 G_0 &= \frac{T_0}{6R_0} \frac{(n^3-n) D_0 } {n D_1 - D_2}, \label{G0} \\
 G_1 &= \frac{1}{6R_1} \frac{(n^3-n) D_3} {n D_4 - D_2},  \label{G1} \\
 D_0 &= (1+\mu_1) + 2\mu_1T_0^{-1}  \left(\frac{R_0}{R_1}\right)^{n+2}- (1-\mu_1) \left(\frac{R_0}{R_1}\right)^{2n}, \label{D0} \\
 D_1 &= (1+\mu_1)(\mu_1 - \mu_i) + 2\mu_1(1-\mu_1) \left(\frac{R_0}{R_1}\right)^{n+1}- (1-\mu_1) (\mu_1-\mu_i) \left(\frac{R_0}{R_1}\right)^{2n}, \label{D1} \\
 D_2 &=  (1+\mu_1)(\mu_1 + \mu_i) + (1-\mu_1) (\mu_1-\mu_i) \left(\frac{R_0}{R_1}\right)^{2n},  \label{D2}  \\
 D_3 &= (\mu_1 + \mu_i)+  2 \mu_1  T_0 \left(\frac{R_0}{R_1}\right)^{n-2} + (\mu_1 - \mu_i) \left(\frac{R_0}{R_1}\right)^{2n}, \label{D3} \\
 D_4 &= (1 - \mu_1)(\mu_1 + \mu_i) + 2 \mu_1 (\mu_1 - \mu_i)  \left(\frac{R_0}{R_1}\right)^{n-1} + (1 - \mu_1)(\mu_1 - \mu_i) \left(\frac{R_0}{R_1}\right)^{2n}. \label{D4}
\end{align}
Therefore, the maximum injection rate for which a disturbance with wave number $n$ is stable, $Q_M(n)$, is bounded below by the right-hand side of equation \eqref{3-LayerUB}:
\begin{equation}\label{3-LayerLB-n}
 Q_M(n) \geq \min\{G_0(n),G_1(n)\}.
\end{equation} 
Considering all wave numbers, the maximum injection rate, $Q_M$, for which the flow is stable satisfies
\begin{equation}\label{3-LayerLB}
 Q_M \geq \min_{n \in \mathbb{N}} \left\{ \min\{G_0(n),G_1(n)\} \right\}.
\end{equation} 
Note that the only terms in the expressions for $G_0$ and $G_1$ that are time-dependent are $R_0$ and $R_1$. As time increases, the interfaces come closer to each other and $R_0/R_1 \to 1$ as $t \to \infty$. Therefore, from equations \eqref{G0} and \eqref{G1}, $G_0 \propto 1/R_0$ and $G_1 \propto 1/R_1$ as $t \to \infty$. This is precisely the relationship between $Q$ and $R$ in equation \eqref{MaxStableQ_2Layer}. Therefore, if the injection rate is chosen such that $Q = G_0$ or $Q = G_1$ then $Q \propto t^{-1/3}$ for $t >> 1$.

\subsubsection{Limiting Cases}\label{sec:LimitingCases}
We now investigate the condition \eqref{3-LayerUB} in the limit when the intermediate layer is very thin ($R_0/R_1 \to 1$). Note that for any three-layer Hele-Shaw flow, the distance between the interfaces decreases with time. Therefore, even if the interfaces are initially far apart, the intermediate layer will eventually become thin.
In the limit as $R_0/R_1 \to 1$, equation \eqref{G0} becomes
\begin{equation}
 \lim_{\frac{R_0}{R_1} \to 1} G_0(n) = \frac{1}{6R_0} \frac{(n^3-n) \left(T_0 + 1\right)}{ n (1  - \mu_i) - ( 1 + \mu_i) }.
\end{equation}
Likewise,
\begin{equation}
 \lim_{\frac{R_0}{R_1} \to 1} G_1(n) = \frac{1}{6R_1} \frac{(n^3-n) \left(T_0 + 1\right)}{ n (1  - \mu_i) - ( 1 + \mu_i) }.
\end{equation}
Since we are considering the limit as $R_0/R_1 \to 1$, it is also true that $R_0 \to R_1$. If we denote $R := R_1$, then 
\begin{equation*}
  \lim_{\frac{R_0}{R_1} \to 1} G_0(n) = \lim_{\frac{R_0}{R_1} \to 1} G_1(n) = \frac{1}{6R} \frac{(n^3-n) \left(T_0 + 1\right)}{ n (1  - \mu_i) - ( 1 + \mu_i) },
\end{equation*}
and the condition \eqref{3-LayerUB} becomes
\begin{equation}\label{ThinLimit}
 Q \leq \frac{1}{6R} \frac{(n^3-n) \left(T_0 + 1\right)}{ n (1  - \mu_i) - ( 1 + \mu_i) }.
\end{equation}
After some algebraic manipulation, this condition becomes
\begin{equation}
 \frac{Qn}{2 R^2} \frac{1  - \mu_i}{1 + \mu_i} - \frac{Q}{2 R^2}   - \frac{(T_0 + 1)}{1 + \mu_i} \frac{(n^3-n)}{12R^3} \leq 0.
\end{equation}
The term on the left-hand side is precisely the two-layer growth rate for a single interface with interfacial tension $T_0 + 1$ (in dimensional variables this is $T_0 + T_1$, the sum of the interfacial tensions of the two interfaces). This is the same thin-layer limit that was found for the exact three-layer growth rate in \cite{gin-daripa:hs-rad}.

Using a similar analysis of the thick-layer limit ($R_0 << R_1$) for $n > 2$, a disturbance with wave number $n$ is stable if the injection rate $Q$ is such that
\begin{equation}\label{thickInner}
 \frac{Qn}{2 R_0^2} \frac{\mu_1  - \mu_i}{\mu_1 + \mu_i} - \frac{Q}{2 R_0^2}   - \frac{T_0}{\mu_1 + \mu_i} \frac{(n^3-n)}{12 R_0^3} \leq 0,
\end{equation}
and
\begin{equation}\label{thickOuter}
 \frac{Qn}{2 R_1^2} \frac{1  - \mu_1}{1 + \mu_1} - \frac{Q}{2 R_1^2}   - \frac{1}{1 + \mu_1} \frac{(n^3-n)}{12 R_1^3} \leq 0.
\end{equation}
Equation \eqref{thickInner} is the condition that the inner interface is stable according to its two-layer growth rate, and equation \eqref{thickOuter} is the condition that the outer interface is stable according to its two-layer growth rate. Therefore, as expected, in the limit of a thick intermediate layer the interfaces are decoupled and the flow is stable if each interface is individually stable.

\subsubsection{Multi-layer Flow}
We now find sufficient conditions on the injection rate to stabilize a flow with an arbitrary number of fluid layers. The approach used for three-layer flows in \S\ref{sec:MaxQ_3Layer} which uses Gershgorin's circle theorem can be adapted to flows with four or more layers by calculating 
the corresponding matrix $\mathbf{M}_N$. However, in order to avoid inverting an $(N+1) \times (N+1)$ matrix, we adopt a different approach. In \cite[p.22]{gin-daripa:hs-rad}, upper bounds are found on the real part of the growth rate for flows with $N$ internal layers. This upper bound is the maximum of $N+1$ expressions, each of which has terms that pertain to the parameter values at one of the interfaces. 
From examining the upper bound, it can be seen that a disturbance with wave number $n$ will be stable if $E_j \leq 0$ for $j = 0,1,...,N$. Using a dimensionless version of the upper bounds in \cite{gin-daripa:hs-rad}, the relationship between $E_j$ and $F_j$ (see equations \eqref{multi:F0}, \eqref{multi:F1}, and \eqref{multi:Fj}) is
\begin{equation}
 \begin{split}
  E_0 = nR_0^2F_0 - \frac{Qn}{2} \mu_i, \quad E_j = nR_j^2F_j, \quad \text{for } j=1,...,N-1, \quad E_N = nR_N^2F_N - \frac{Qn}{2}.
 \end{split}
\end{equation}
$E_j \leq 0$ for all $j$ if
\begin{equation}\label{MaxStableQ_NLayer2}
 Q \leq \min \left\{ \frac{1}{6R_0} \frac{T_0 n (n^2-1)}{n(\mu_1-\mu_i) - \mu_i}, \min_{j=1,...,N-1} \frac{1}{6R_j} \frac{T_j (n^2-1)}{\mu_{j+1}-\mu_{j}},  \frac{1}{6R_N} \frac{n (n^2-1)}{n(1-\mu_N) - 1} \right\}.
\end{equation}
Therefore, a lower bound on the maximum stable injection rate $Q_M$ is given by
\small
\begin{equation}\label{MaxStableQ_NLayer}
 Q_M \geq \min_{n \in \mathbb{N}} \left\{ \min \left\{ \frac{1}{6R_0} \frac{T_0 n (n^2-1)}{n(\mu_1-\mu_i) - \mu_i}, \min_{j=1,...,N-1} \frac{1}{6R_j} \frac{T_j (n^2-1)}{\mu_{j+1}-\mu_{j}},  \frac{1}{6R_N} \frac{n (n^2-1)}{n(1-\mu_N) - 1} \right\} \right\}.
\end{equation}
\normalsize
Notice the similarity between the terms in \eqref{MaxStableQ_NLayer} and the expression for two-layer flows given by equation \eqref{MaxStableQ_2Layer}.

The above approach can be used to obtain a bound on a stable injection rate for three-layer flows. However, the approach taken in \S\ref{sec:MaxQ_3Layer} using Gershgorin's Circle theorem generally produces sharper bounds. This is because the effects of the interfaces have been decoupled in equation \eqref{MaxStableQ_NLayer} and in the decoupling process something is lost.

\subsection{Choosing $Q(t)$ for a time-independent eigenvector of $\mathbf{M}_N$}\label{Qt_for_F0}
Note that the matrix $\mathbf{M}_1$ for three-layer flow is time-dependent (see equation \eqref{M_3Layer}). Therefore, the eigenvalues and eigenvectors evolve in time. However, it is possible to have an eigenvector which persists for all time if a time-dependent injection rate is 
chosen cleverly. Recall from equation \eqref{multi:F0} the expression for $F_0$ which the matrix $\mathbf{M}_1$ in \eqref{M_3Layer} depends on:
\begin{equation*}
 F_0 = \frac{Q n}{2 R_0^2} (\mu_1 - \mu_i) - \frac{T_0}{12} \frac{n^3 - n}{R_0^3}.
\end{equation*}
If 
\begin{equation}\label{Q_F0}
 Q(t) = \frac{T_0 (n^2-1)}{6R_0(t) (\mu_1-\mu_i)},
\end{equation}
then $F_0 = 0$ for all $t$. This results in an injection rate of the form $Q(t) \propto t^{-1/3}$. If the injection rate given by \eqref{Q_F0} is chosen, then the matrix $\textbf{M}_1$ in \eqref{M_3Layer} becomes
\begin{equation*}
\mathbf{M}_1 = \left( 
\begin{array}{cc}
  - \frac{Q}{2 R_0^2} &
 \frac{2 \mu_1 \left(\frac{R_0}{R_1}\right)^{n-1} F_1}{(\mu_1 - \mu_i) (1 - \mu_1) \left(\frac{R_0}{R_1}\right)^{2n} + (\mu_1 + \mu_i) (1 + \mu_1)} \\
 0 & 
 \frac{\left\{(\mu_1 + \mu_i) + (\mu_1 - \mu_i) \left(\frac{R_0}{R_1}\right)^{2n} \right\}  F_1 }{(\mu_1 - \mu_i) (1 - \mu_1) \left(\frac{R_0}{R_1}\right)^{2n} + (\mu_1 + \mu_i) (1 + \mu_1)} - \frac{Q}{2 R_1^2} \\
\end{array} \right).
\end{equation*}
Therefore, $\begin{pmatrix} 1  & 0 \end{pmatrix}^T$ is an eigenvector with an eigenvalue of $-Q/(2 R_0^2)$. If the initial condition is some constant multiple of this eigenvector (i.e. only the inner interface has a disturbance of wave number $n$), then the outer interface will stay circular and the disturbance on the inner interface will decay. This is true even if the 
outer interface individually is very unstable. Similarly, if the injection rate is chosen so that $F_1 = 0$, then $\begin{pmatrix} 0  & 1 \end{pmatrix}^T$ is an eigenvector with an eigenvalue of $-Q/(2  R_1^2)$. An initial condition in which only the outer interface has a disturbance of wave number $n$ will result in the disturbance decaying and the inner interface remaining circular.

Note that this result can also be extended to flows with an arbitrary number of fluid layers. Recall that the expression for the matrix $\mathbf{M}_N$ is given by equation \eqref{M}:
\begin{equation*}
 \textbf{M}_N = \mathbf{\widetilde{M}}_N^{-1} \left( \begin{array}{ccc} F_0 &  \dots & 0 \\ \vdots & \ddots & \vdots \\ 0 & \dots & F_N \end{array} \right) - \frac{Q}{2} \left( \begin{array}{ccc} \frac{1}{R_0^2} &  \dots & 0 \\ \vdots & \ddots & \vdots \\ 0 & \dots & \frac{1}{R_N^2} \end{array} \right).
\end{equation*}
The formula for $F_0$ is the same as that for three-layer flow. Therefore, by using the injection rate given by equation \eqref{Q_F0}, $F_0 = 0$ for all $t$. In this case, even without calculating the matrix $\widetilde{\mathbf{M}}_N^{-1}$, it is clear that $\begin{pmatrix} 1  & 0 & \cdots & 0 \end{pmatrix}^T$ is an eigenvector of $\mathbf{M}_N$ with eigenvalue $-Q/(2 R_0^2)$.

Similarly, we can use the expression \eqref{multi:Fj} for $F_j$ to find that using the injection rate
\begin{equation}\label{Q_Fj}
 Q(t) = \frac{T_j (n^2-1)}{6R_j(t) (\mu_{j+1}-\mu_j)}
\end{equation}
will result in $F_j = 0$ for all $t$. There will be an eigenvector with an entry 1 in the $j$th position and zeros everywhere else, and the corresponding eigenvalue is $-Q/(2 R_j^2)$.

Note that in all of these cases, the time-independent eigenvector has a negative eigenvalue. Therefore, the span of the eigenvector is a stable manifold of the system. If the initial condition is such that only one interface is perturbed and the perturbation consists of only a single wave number and the injection rate is as prescribed by equation \eqref{Q_Fj}, the disturbance will decay. This is independent of the values of all of the parameters including viscosity and interfacial tension and is summarized in the following theorem.

\begin{theorem}
In any multi-layer radial Hele-Shaw flow, if all of the interfaces are circular except for one interface which is perturbed with a monochromatic wave, then there exists a time-dependent injection rate such that the circular interfaces remain circular and the disturbance on the perturbed interface decays.
\end{theorem}

\section{Numerical Results}\label{sec:Numerical}
In this section, we numerically integrate the dynamical system \eqref{MultiLayerMatEqn} to compute the motion of the interfaces within the linear theory and calculate the growth rates of interfacial disturbances. This will reveal some interesting dynamics of interfaces in multi-layer flows. We should mention right at the outset that in some figures (see Figures \ref{fig:InnerDisturb_Interfaces_t0}, \ref{fig:InnerDisturb_Interfaces_t30}, and \ref{fig:F0_0_Interfaces_t300}) which depict the interfaces, the disturbances grow to a point that we would expect to be beyond the limits of the linear theory. However, because of the linearity of the problem, reducing the amplitude of the disturbances would not change the growth rate or the behavior of the interfaces. Therefore, we have chosen larger disturbances so that the reader can more clearly see the patterns of the disturbances on the interfaces.

\subsection{Constant Injection Rate}\label{sec:Numerical_Constant}
We start by considering flows with a constant injection rate. The motion of the interfaces is computed using the equations derived in \S\ref{sec:Preliminaries}. Much of the interesting interfacial dynamics in $N$-layer flows also appear in three-layer and four-layer flows so we only show results in these two cases. The values of the parameters are given in the various figure captions below. Many of the figures below show trajectories in phase space to depict the evolution of the amplitudes of a disturbance with wave number $n$ for the inner interface ($A_n$) versus the outer interface ($B_n$) for three-layer flow or the inner interface ($A_n$), intermediate interface ($B_n$), and outer interface ($C_n$) for four-layer flow. The `*' denotes the amplitudes at time $t = 0$. The markers on the line are equally spaced in time. The diagonal $A_n = B_n$ is given for reference.

Figure \ref{fig:InnerDisturb_phase} shows the growth of interfacial disturbances for a three-layer flow in which initially only the inner interface has a disturbance ($A_n \neq 0, B_n = 0$). According to the system of equations \eqref{MultiLayerMatEqn} with $N=1$ and \eqref{M_3Layer}, the ODEs governing the interfaces are coupled. Therefore, we expect that the instability of the inner interface will be transferred to the outer interface at time $t > 0$. The trajectory in Figure \ref{fig:InnerDisturb_phase} matches this expectation because a disturbance immediately forms on the outer interface as the disturbance of the inner interface initially decays. An interesting thing to note is that as the disturbance develops on the outer interface ($B_n$), its sign is negative. Since the amplitude of the disturbance on the inner interface ($A_n$) is positive, this corresponds to disturbances that are out of phase with a phase shift that is half of the period. However, the curve eventually crosses the $B_n$-axis, meaning that the inner interface becomes circular, but then the disturbance of the inner interface begins to grow again but in phase with the disturbance of the outer interface. 
\begin{figure}
\begin{subfigure}[b]{.5\textwidth}
\centering
\includegraphics[width=2.5 in,height=2 in]{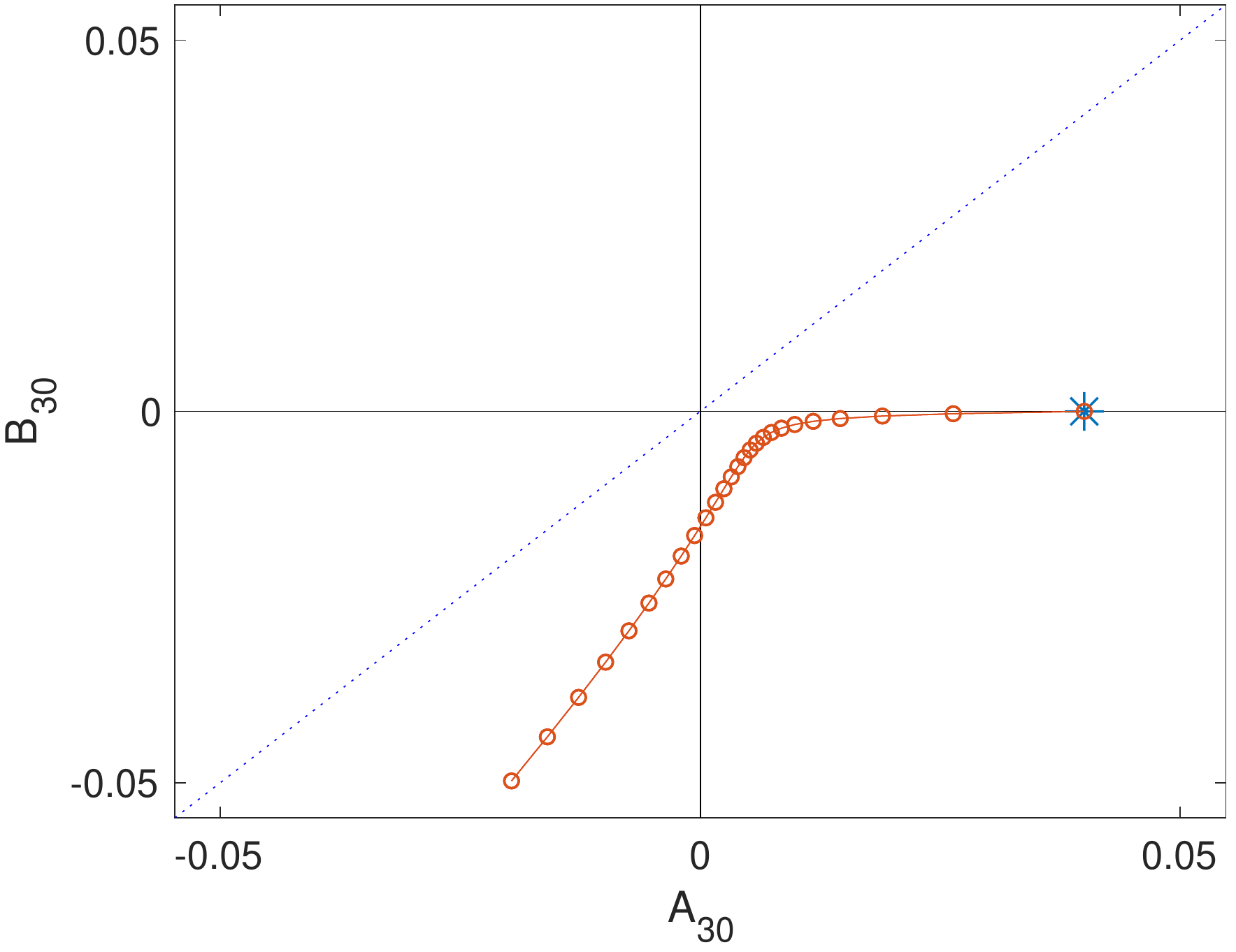}
\caption{}
\label{fig:InnerDisturb_phase}
\end{subfigure}
\begin{subfigure}[b]{.5\textwidth}
  \centering
  \includegraphics[width=2.5 in,height=2 in]{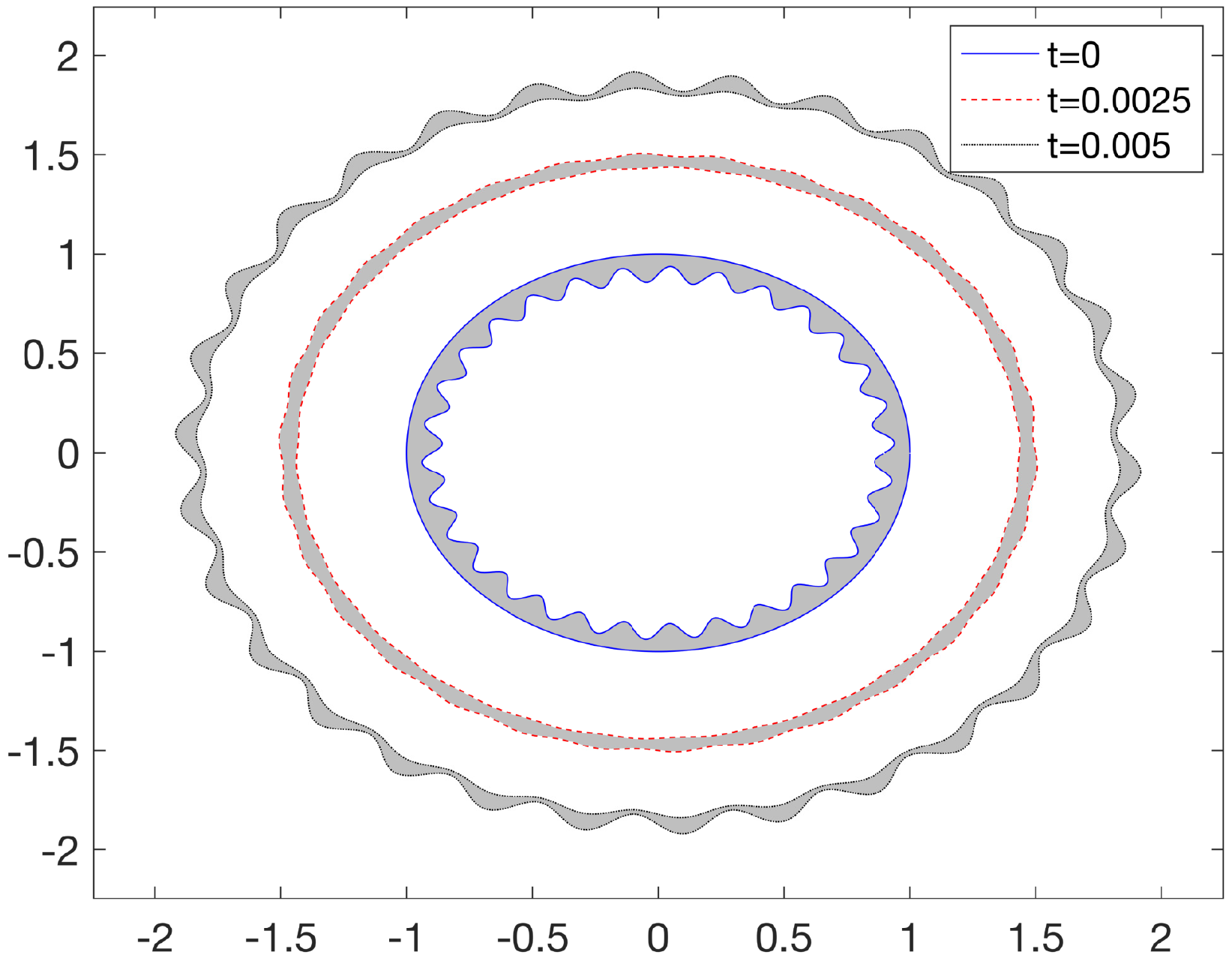}
  \caption{}
\label{fig:InnerDisturb_Interfaces}
\end{subfigure}
\caption{A plot of the interfacial disturbances of three-layer flow in the case that only the inner interface is disturbed initially. Plot (a) is the phase plane plot of the amplitude of the disturbance on the inner interface ($A_{30}$) versus the amplitude of the disturbance on the outer interface ($B_{30}$). Plot (b) shows the interfaces at times $t = 0$, $t = 0.0025$, and $t = 0.005$. The parameter values are $Q = 500, \mu_i = 0.2, \mu_1 = 0.4, \mu_o = 1, T_0 = T_1 = 1, R_0(0) = 0.9, R_1(0) = 1, n = 30$, $A_{30}(0) = 0.04$, $B_{30}(0) = 0, 0 \leq t \leq 0.005$.}
\label{fig:InnerDisturb}
\end{figure}

In order to visualize the behavior of the interfaces, the interfaces corresponding to Figure \ref{fig:InnerDisturb_phase} are plotted at three different times in Figure \ref{fig:InnerDisturb_Interfaces}. The intermediate fluid that lies between the two interfaces is shaded grey. Because the disturbances are small compared to the radii of the unperturbed interfaces, zoomed in plots of the interfaces at each time step are given in Figure \ref{fig:InnerDisturb_Interfaces_zoom}.
Note that in Figure \ref{fig:InnerDisturb_Interfaces_t0}, the outer interface is circular and only the inner interface has a disturbance. Figure \ref{fig:InnerDisturb_Interfaces_t15} shows that the disturbances are out of phase at time $t = 0.0025$. However, at time $t = 0.005$, the disturbances are in phase as illustrated in Figure \ref{fig:InnerDisturb_Interfaces_t30}. Also, the outer interface which initially had no disturbance now has a larger disturbance than the inner interface.

\begin{figure}
\begin{subfigure}[b]{.33\textwidth}
  \centering
  \includegraphics[width=1.7 in,height=1.7 in]{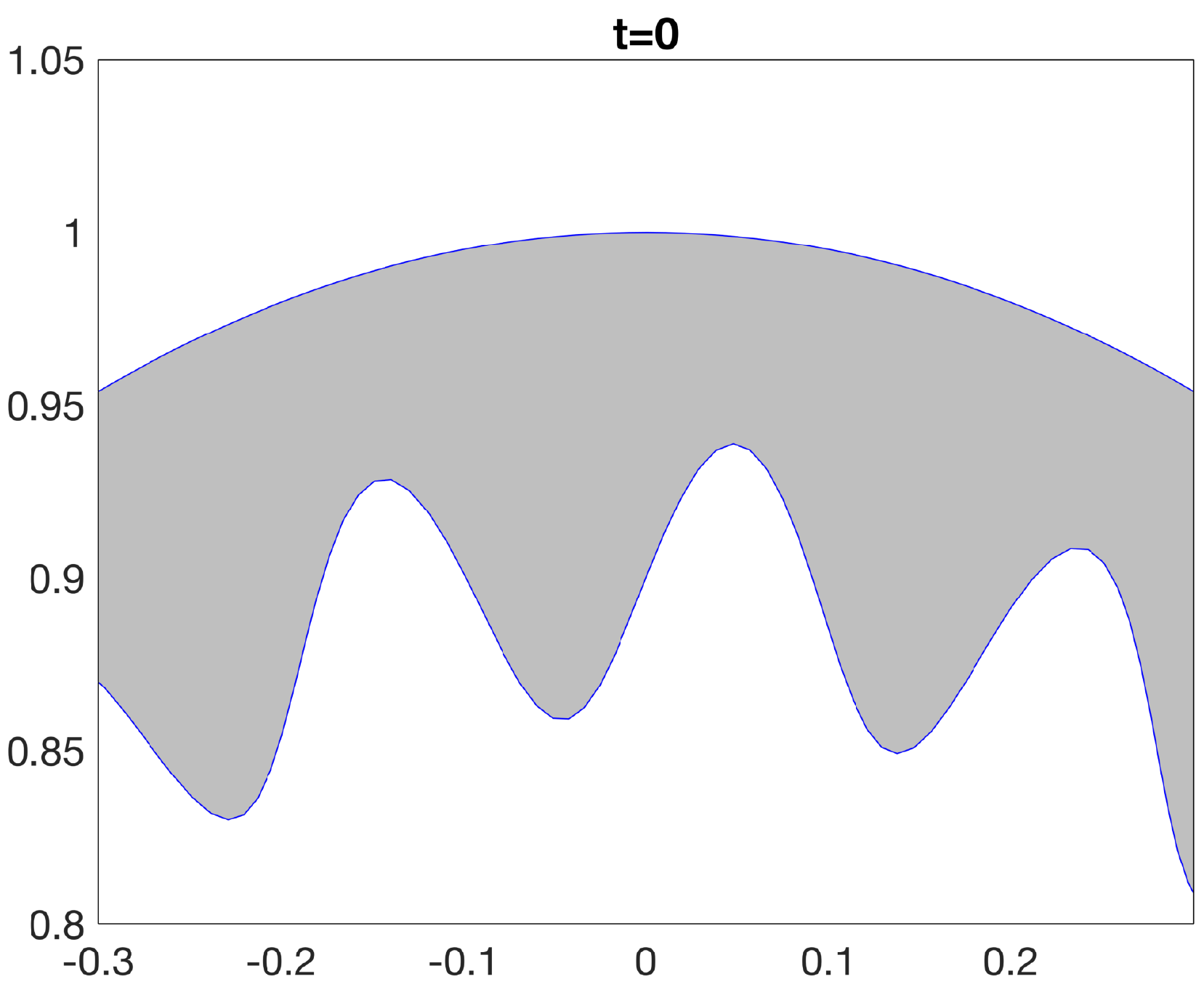}
  \caption{}
  \label{fig:InnerDisturb_Interfaces_t0}
\end{subfigure}%
\begin{subfigure}[b]{.33\textwidth}
  \centering
  \includegraphics[width=1.7 in,height=1.7 in]{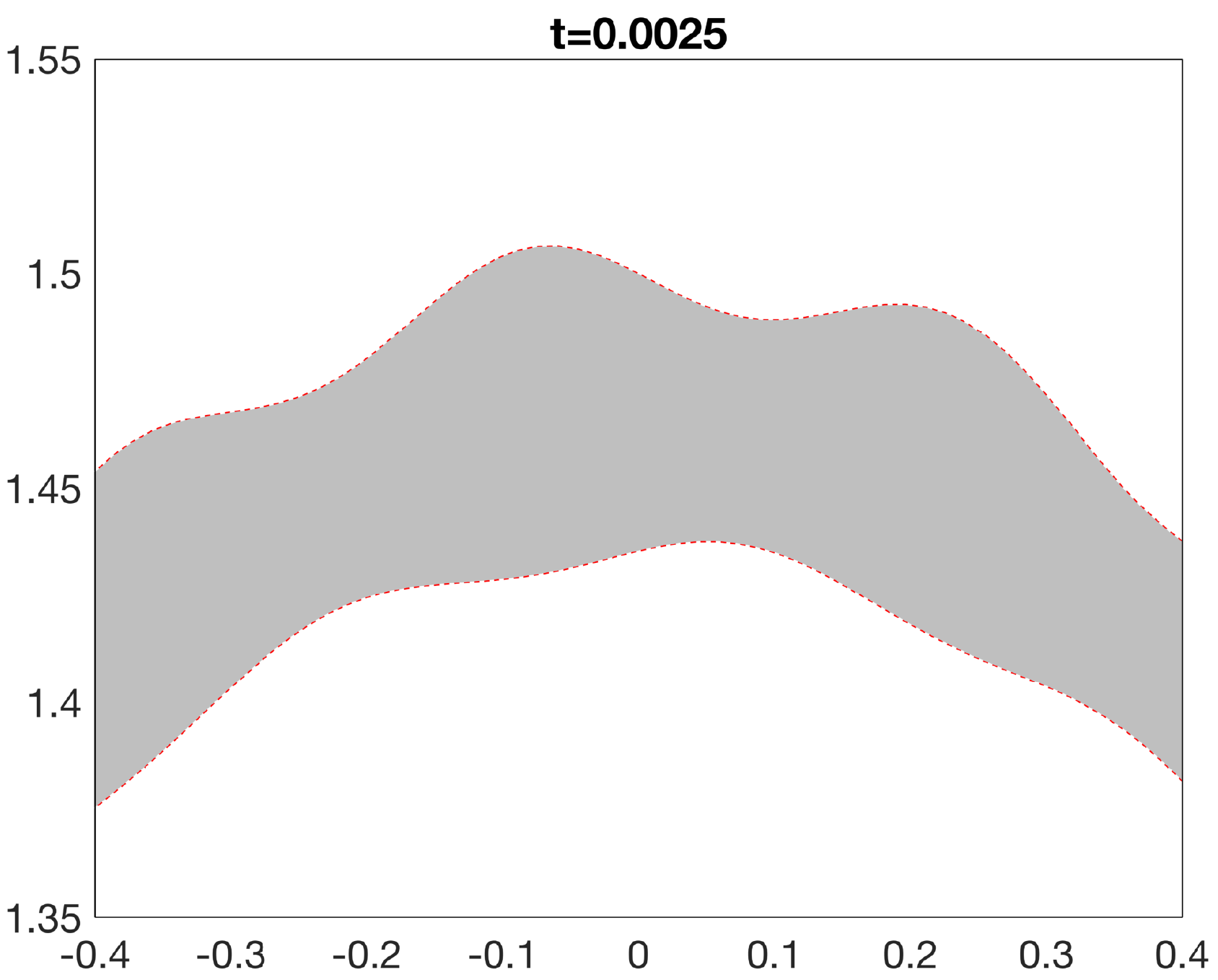}
  \caption{}
  \label{fig:InnerDisturb_Interfaces_t15}
\end{subfigure}
\begin{subfigure}[b]{.33\textwidth}
  \centering
  \includegraphics[width=1.7 in,height=1.7 in]{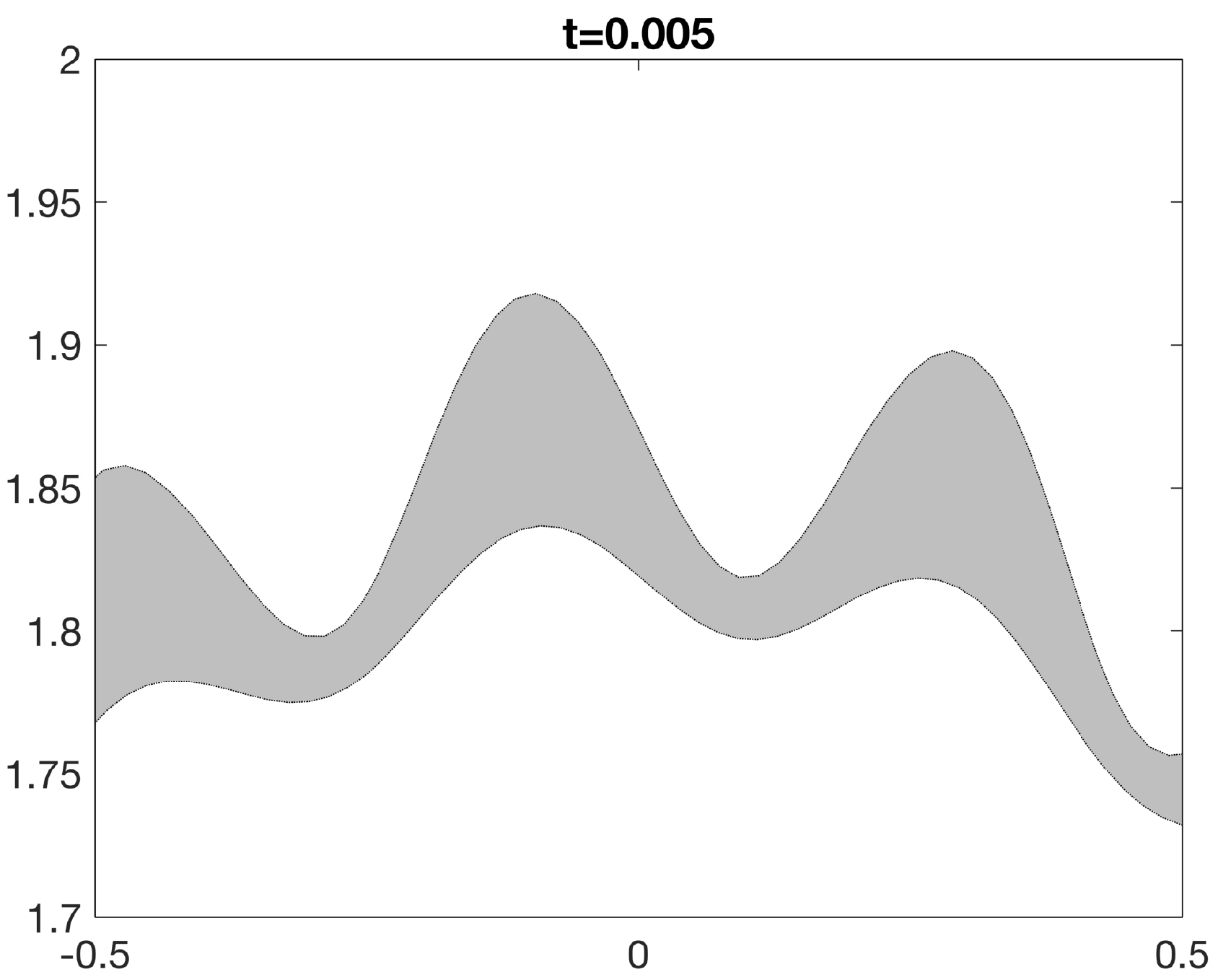}
  \caption{}
  \label{fig:InnerDisturb_Interfaces_t30}
\end{subfigure}
\caption{Plots of the interfaces in the case that only the inner interface is disturbed initially (see Figure \ref{fig:InnerDisturb}). Only a segment of the middle layer is shown at three different time levels in order to show the wave patterns on the interfaces more clearly. The interfaces are shown at times (a) $t = 0$, (b) $t = 0.0025$, and (c) $t = 0.005$.}
\label{fig:InnerDisturb_Interfaces_zoom}
\end{figure}

In the case of four-layer flow, there are even more possibilities -- the disturbances of all three interfaces could be in phase or any one of the three interfaces could be out of phase with the other two. In order to explore these possibilities, we perform some simulations for four-layer flow. We use flows that are analogous to the three-layer flow considered previously. For Figure \ref{fig:InnerDisturb}, there were interfaces initially located at $R_0(0) = 0.9$ and $R_1(0) = 1$, and a disturbance was initially given to only the inner interface. The associated viscous profile is given in Figure \ref{fig:3Layer_viscousprofile}. 
To have an analogous four-layer flow, there are three options. An interface can be added inside of the inner interface, outside of the outer interface, or between the two interfaces. The first two of those cases are considered and the initial viscous profiles are given in Figures \ref{fig:4Layer_viscousprofile_inner} and \ref{fig:4Layer_viscousprofile_outer}. Figure \ref{fig:4Layer_viscousprofile_inner} corresponds to adding an interface inside of the inner interface at $R = 0.8$. Figure \ref{fig:4Layer_viscousprofile_outer} corresponds to adding an interface outside of the outer interface, but note that this changes the scaling so that the new outermost interface is initially at $R = 1$.
\begin{figure}
\begin{subfigure}[b]{.33\textwidth}
  \centering
  \includegraphics[width=1.7 in,height=1.7 in]{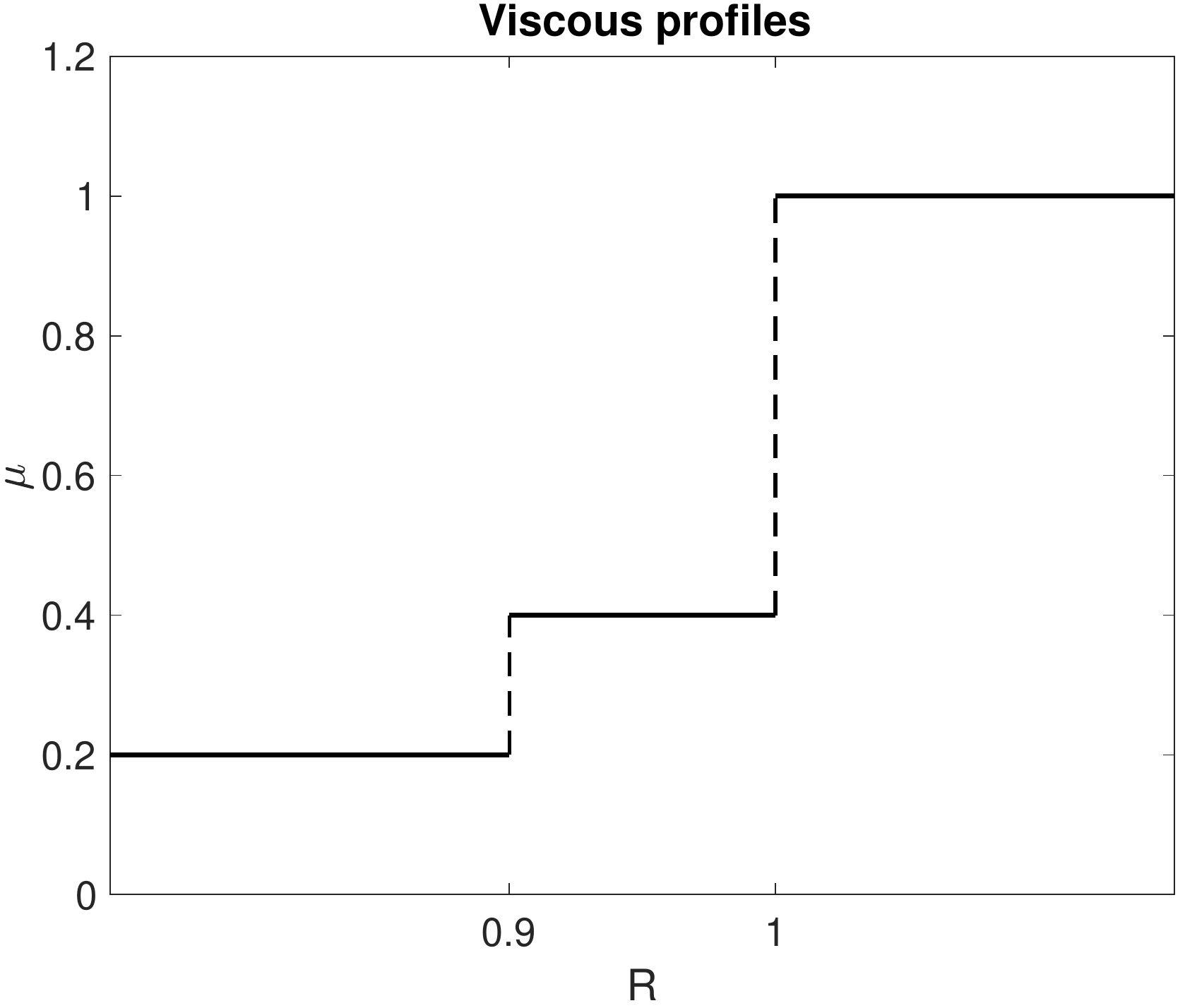}
  \caption{}
  \label{fig:3Layer_viscousprofile}
\end{subfigure}%
\begin{subfigure}[b]{.33\textwidth}
  \centering
  \includegraphics[width=1.7 in,height=1.7in]{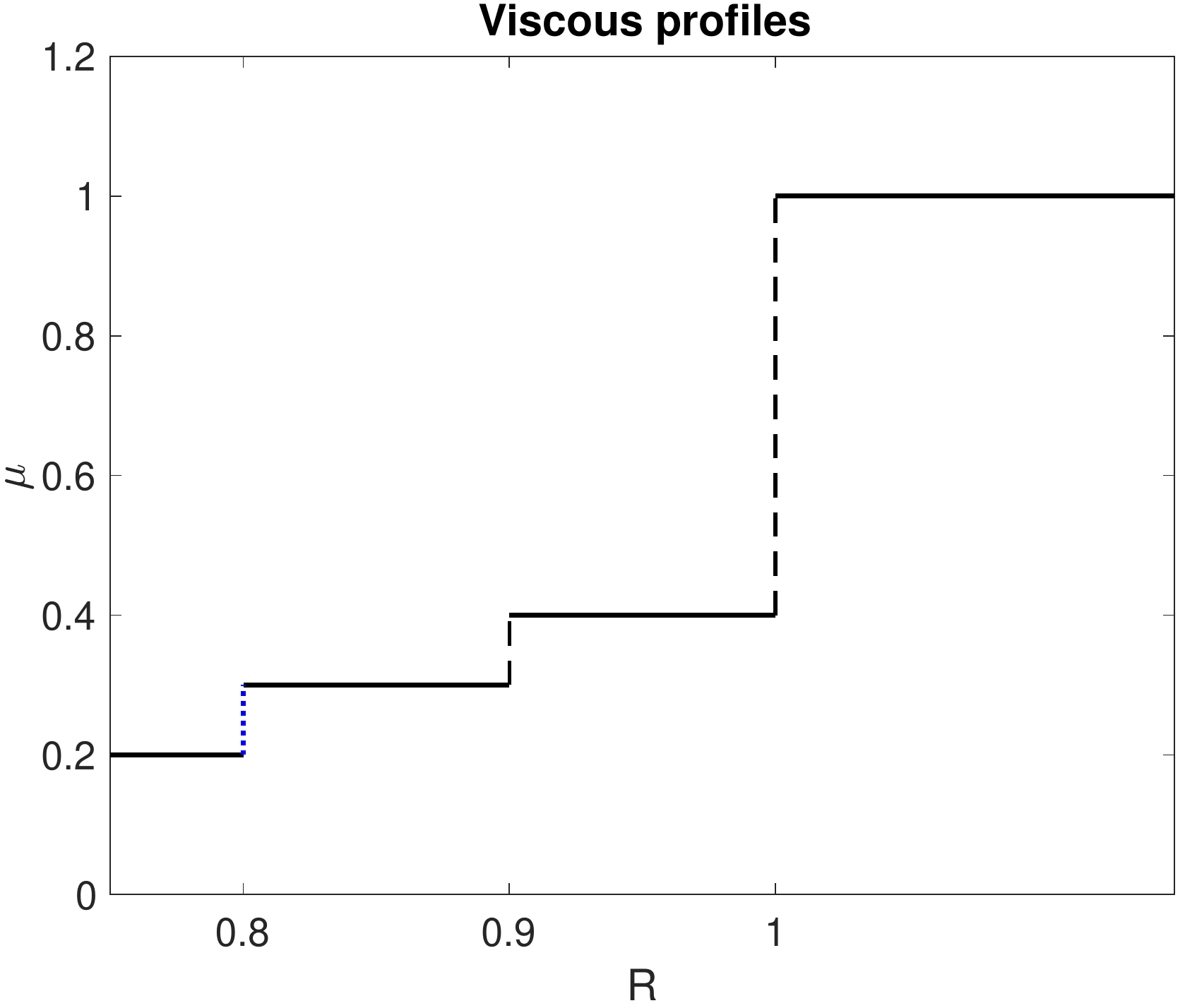}
  \caption{}
  \label{fig:4Layer_viscousprofile_inner}
\end{subfigure}
\begin{subfigure}[b]{.33\textwidth}
  \centering
  \includegraphics[width=1.7 in,height=1.7 in]{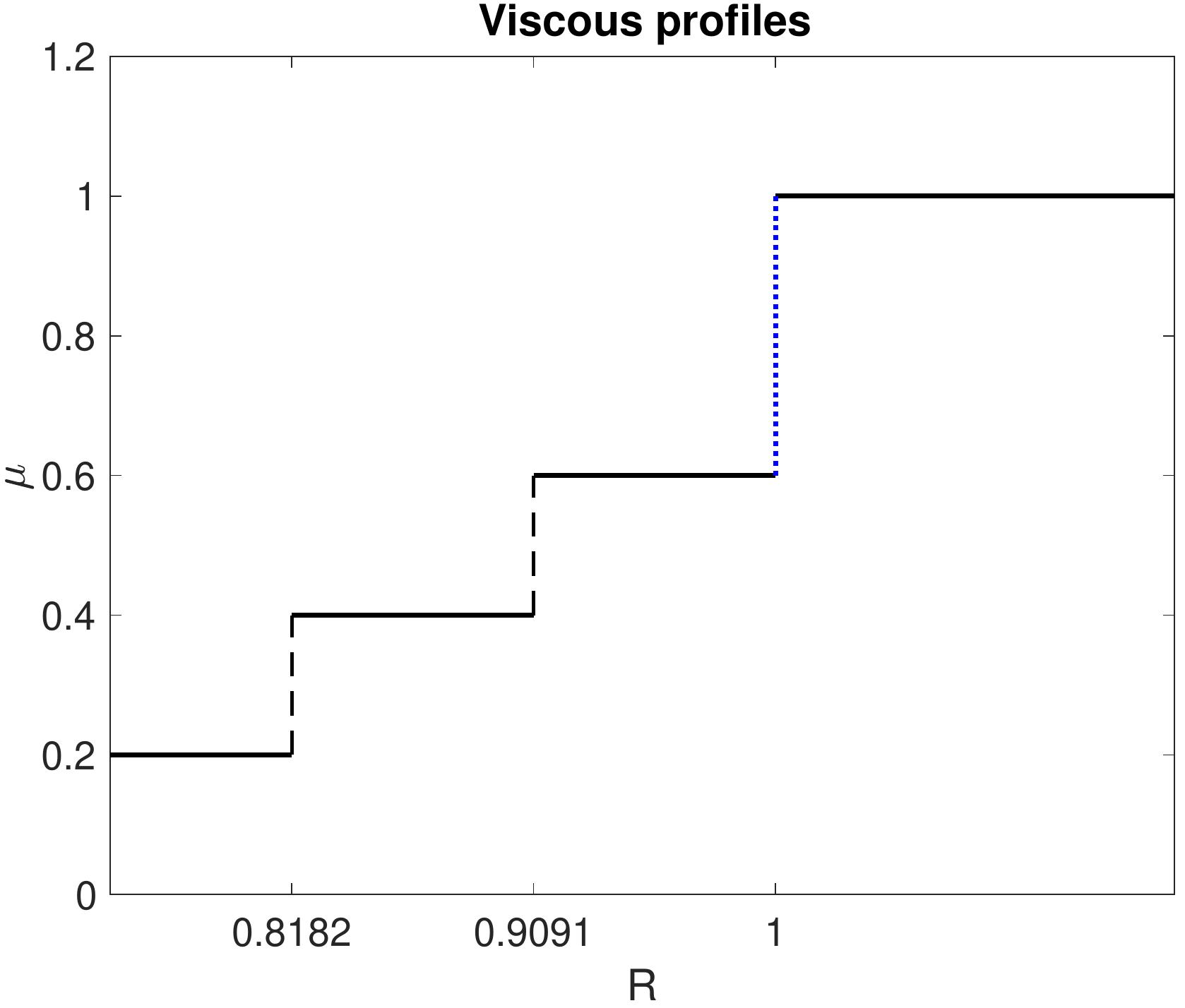}
  \caption{}
  \label{fig:4Layer_viscousprofile_outer}
\end{subfigure}
\caption{Plots of the initial viscous profiles associated with (a) Figure \ref{fig:InnerDisturb}, (b) Figure \ref{fig:4Layer_inner}, and (c) Figure \ref{fig:4Layer_outer}. The four-layer flows (b) and (c) correspond to the three-layer flow in (a) but with one additional interface added. Note that the interfaces located at $R = 0.9$ and $R = 1$ in (a) and (b) have the same physical location as the interfaces located at $R = 9/11$ and $R = 10/11$ in (c), but the values are different because of the scaling.}
\label{fig:4Layer_viscousprofiles}
\end{figure}

Figure \ref{fig:4Layer_inner} shows the evolution of the interfacial disturbances when an interface has been added inside the inner interface of the three-layer flow. The initial viscous profile is shown in Figure \ref{fig:4Layer_viscousprofile_inner}. Similar to the corresponding three-layer flow, only the interface at $R = 0.9$ is initially disturbed. In this case, that is the middle interface ($B_n$). As shown in Figure \ref{fig:4Layer_inner}, disturbances form on the other two interfaces for $t > 0$. Figure \ref{fig:4Layer_inner_BC} shows the amplitudes of the disturbances of the two outermost interfaces. By comparing this with Figure \ref{fig:InnerDisturb_phase}, we see that the qualitative behavior of these two interfaces is unchanged by the presence of a third interface. The disturbances form out of phase and then eventually become in phase. Figure \ref{fig:4Layer_inner_AC} shows that the inner and outer interfaces are always in phase. Therefore, for small positive values of $t$, the outer and inner interfaces are in phase while the middle interface is out of phase. Then, after some time, all three interfaces are in phase.
\begin{figure}
\begin{subfigure}[b]{.33\textwidth}
  \centering
  \includegraphics[width=1.7 in,height=1.7 in]{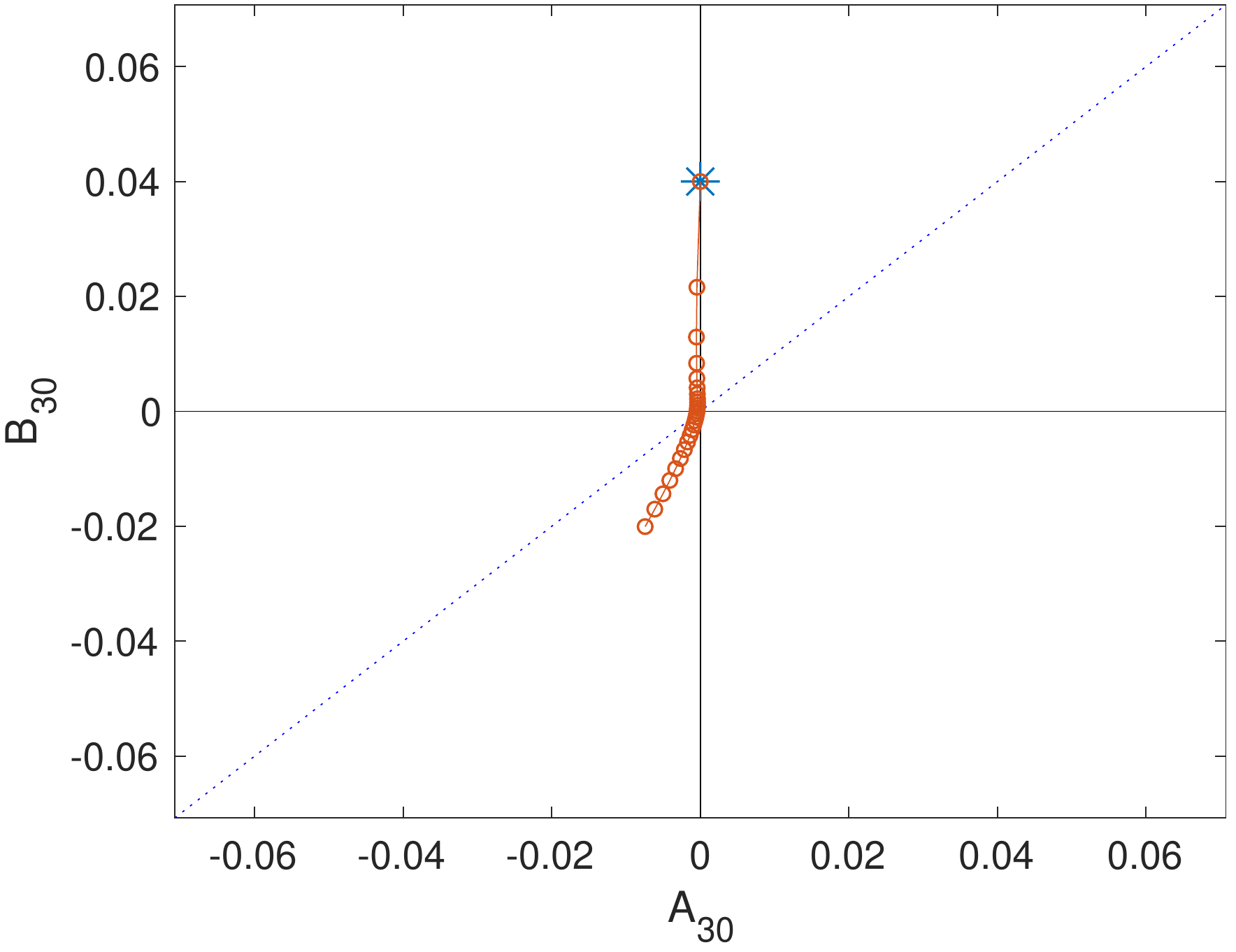}
  \caption{}
  \label{fig:4Layer_inner_AB}
\end{subfigure}%
\begin{subfigure}[b]{.33\textwidth}
  \centering
  \includegraphics[width=1.7 in,height=1.7in]{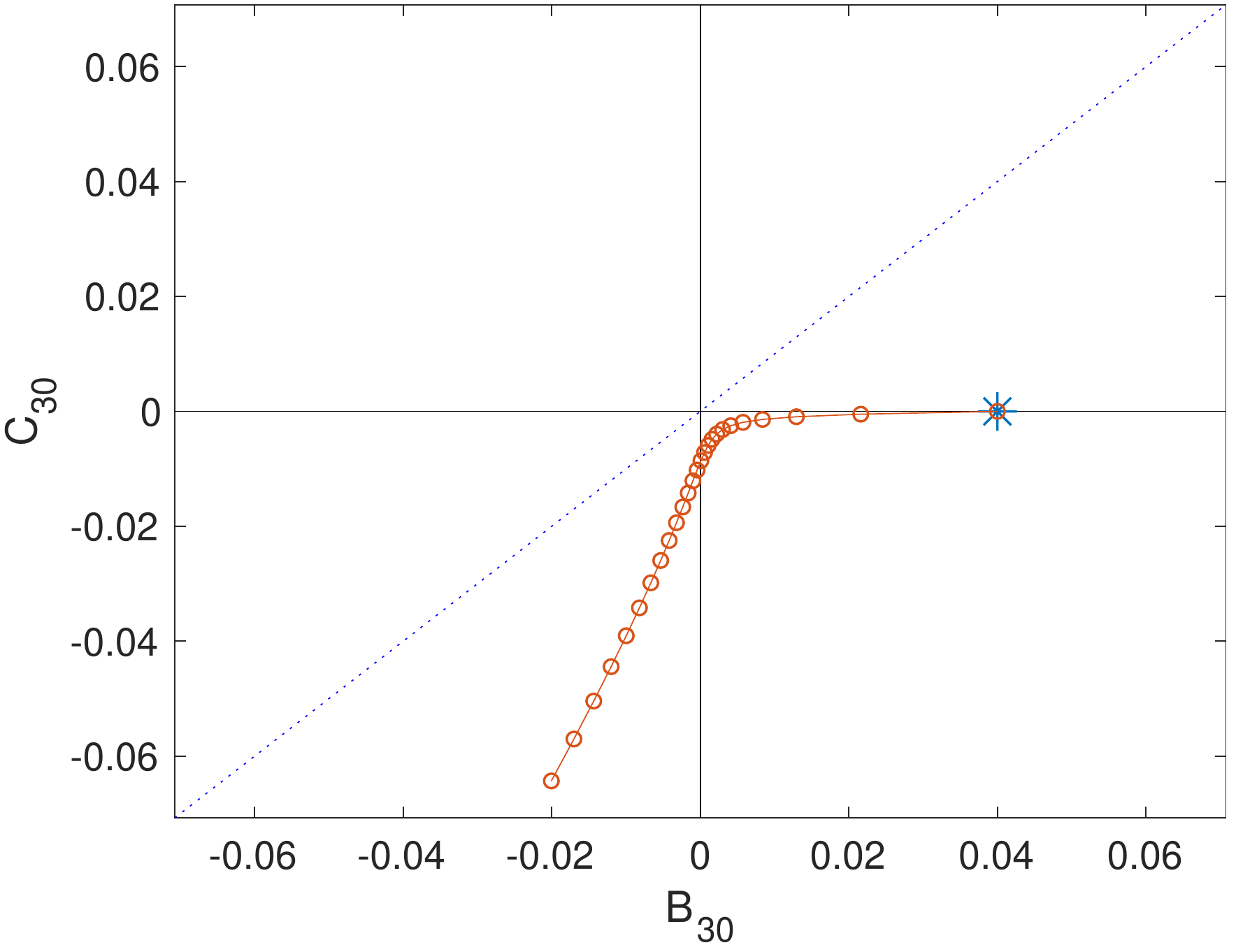}
  \caption{}
  \label{fig:4Layer_inner_BC}
\end{subfigure}
\begin{subfigure}[b]{.33\textwidth}
  \centering
  \includegraphics[width=1.7 in,height=1.7 in]{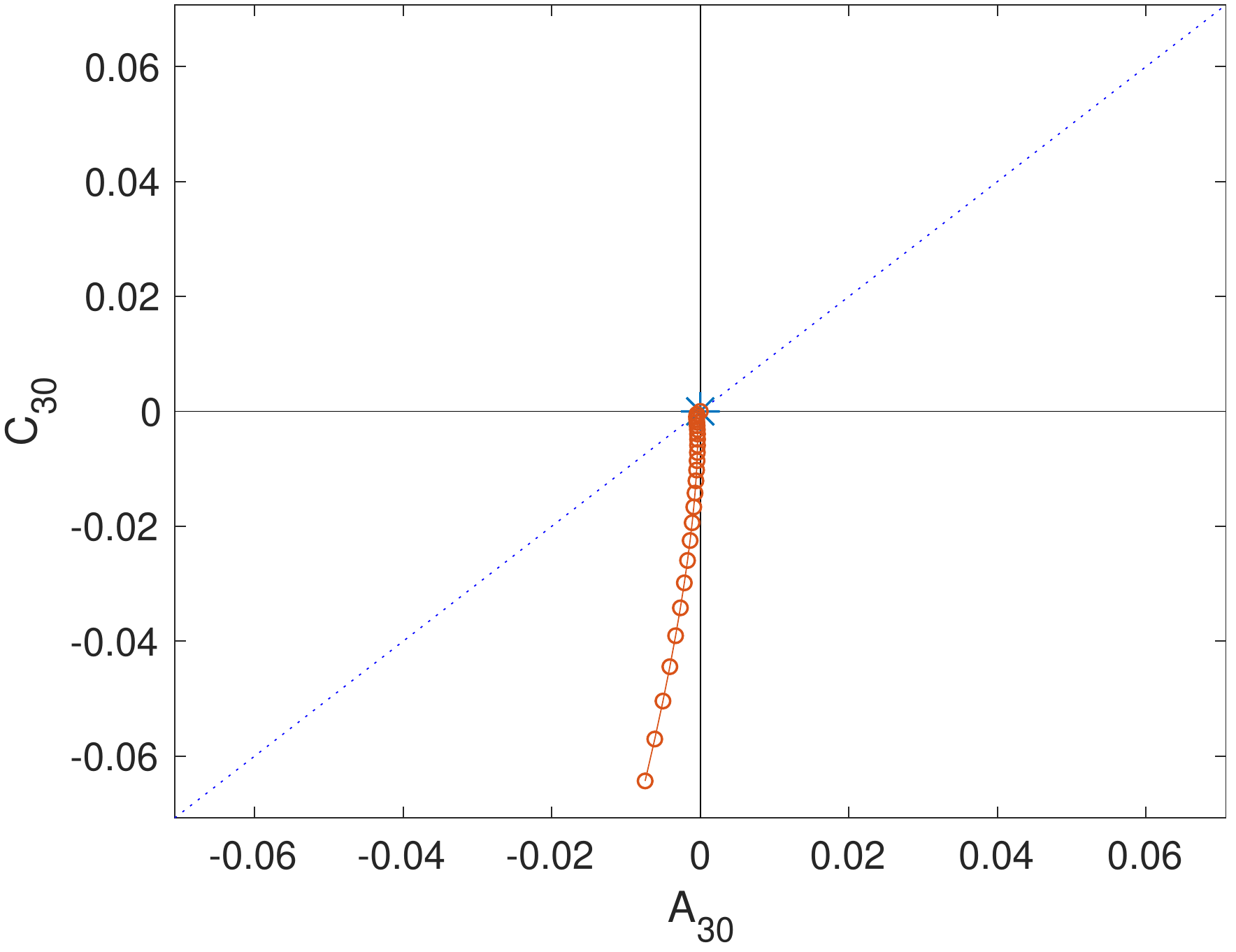}
  \caption{}
  \label{fig:4Layer_inner_AC}
\end{subfigure}
\caption{Plots of the amplitude of interfacial disturbances for four-layer flow in the case that only the middle interface is disturbed initially. Plot (a) shows the amplitude of the disturbance on the inner interface ($A_{30}$) versus the amplitude of the disturbance on the middle interface ($B_{30}$), plot (b) shows the middle interface ($B_{30}$) versus the outer interface ($C_{30}$), and plot (c) shows the inner interface ($A_{30}$) versus the outer interface ($C_{30}$). The parameter values are $Q = 500$, $\mu_i = 0.2$, $\mu_1 = 0.3$, $\mu_2 = 0.4$, $\mu_o = 1$, $T_0 = T_1 = T_2 = 1$, $R_0(0) = 0.8$, $R_1(0) = 0.9$, $R_2(0) = 1$, $n = 30$, $A_{30}(0) = 0$, $B_{30}(0) = 0.04$, $C_{30}(0) = 0$, $0 \leq t \leq 0.005$.}
\label{fig:4Layer_inner}
\end{figure}

Plot (a) is the phase plane plot of the amplitude of the disturbance on the inner interface ($A_{30}$) versus the amplitude of the disturbance on the outer interface ($B_{30}$).

Figure \ref{fig:4Layer_outer} shows the amplitudes of the interfacial disturbances when an interface is added outside of the two interfaces of the corresponding three-layer flow. The initial viscous profile is given by Figure \ref{fig:4Layer_viscousprofile_outer}. Only the innermost interface is perturbed ($A_n$). 
Figure \ref{fig:4Layer_outer_AB} shows the disturbances of the two interfaces which were present in the three-layer flow. They still start out of phase and then move in phase, but now it is the disturbance on the outer of the two interfaces which vanishes at the transition (compare with Figure \ref{fig:InnerDisturb_phase} where $A_n$ vanishes). Even more interesting behavior occurs in Figure \ref{fig:4Layer_outer_BC} which shows the amplitudes of disturbances for the two outermost interfaces. The curve goes through the third quadrant, then the second quadrant, and finally into the first quadrant. That means that the disturbances of the outermost interfaces form in phase, then move out of phase, and then move back in phase. This shows that the addition of a third interface can significantly change the dynamics of the system.
\begin{figure}
\begin{subfigure}[b]{.33\textwidth}
  \centering
  \includegraphics[width=1.7 in,height=1.7 in]{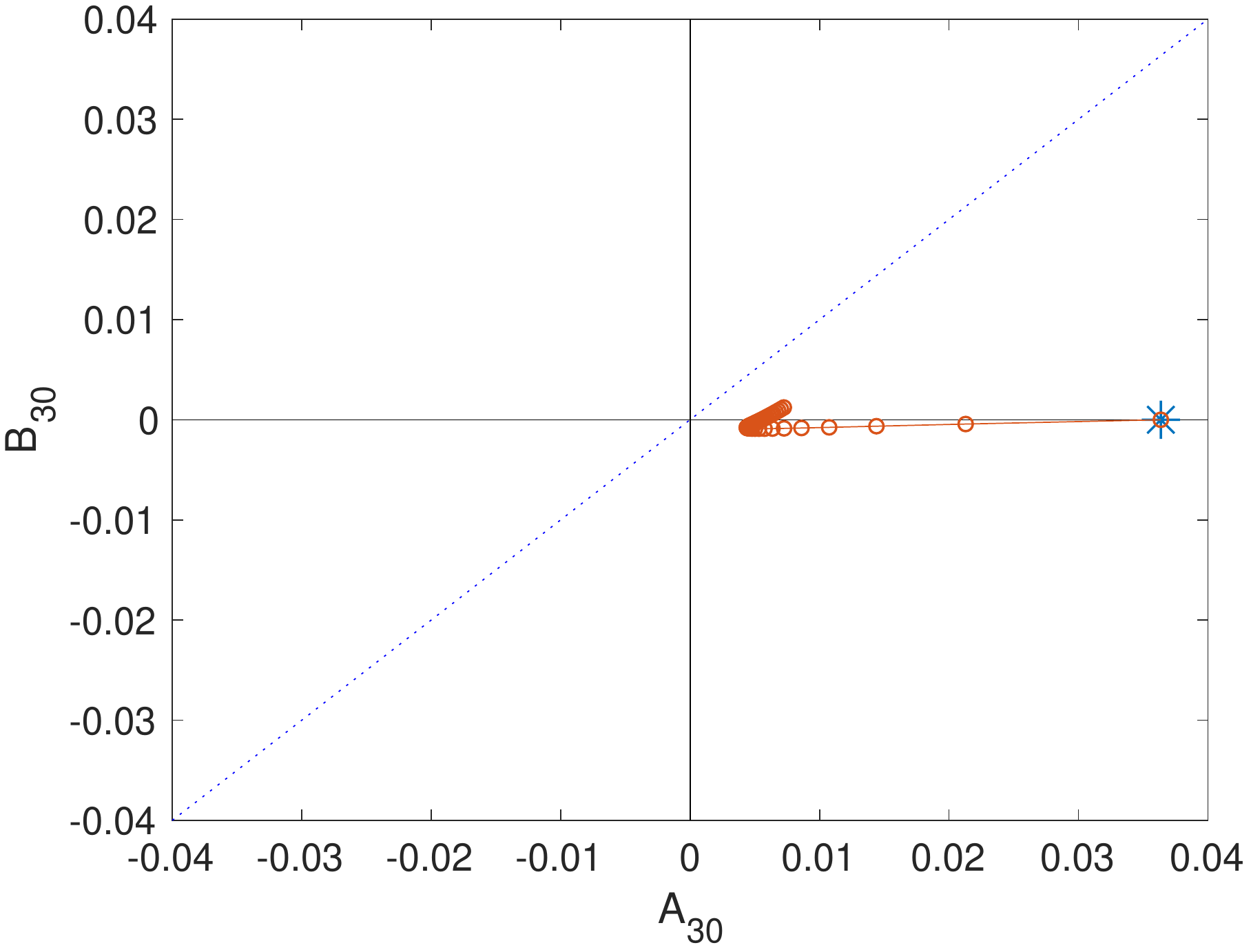}
  \caption{}
  \label{fig:4Layer_outer_AB}
\end{subfigure}%
\begin{subfigure}[b]{.33\textwidth}
  \centering
  \includegraphics[width=1.7 in,height=1.7in]{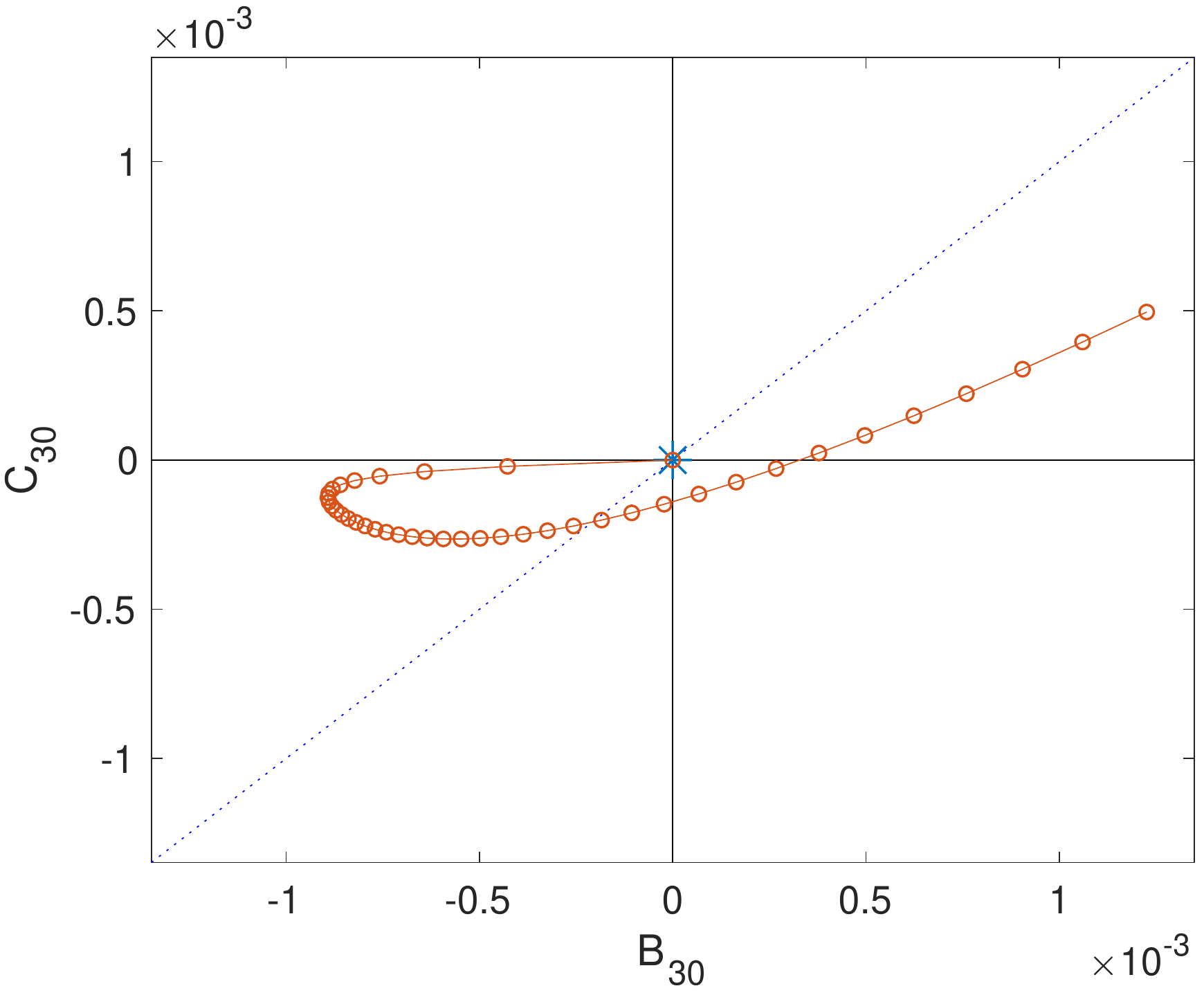}
  \caption{}
  \label{fig:4Layer_outer_BC}
\end{subfigure}
\begin{subfigure}[b]{.33\textwidth}
  \centering
  \includegraphics[width=1.7 in,height=1.7 in]{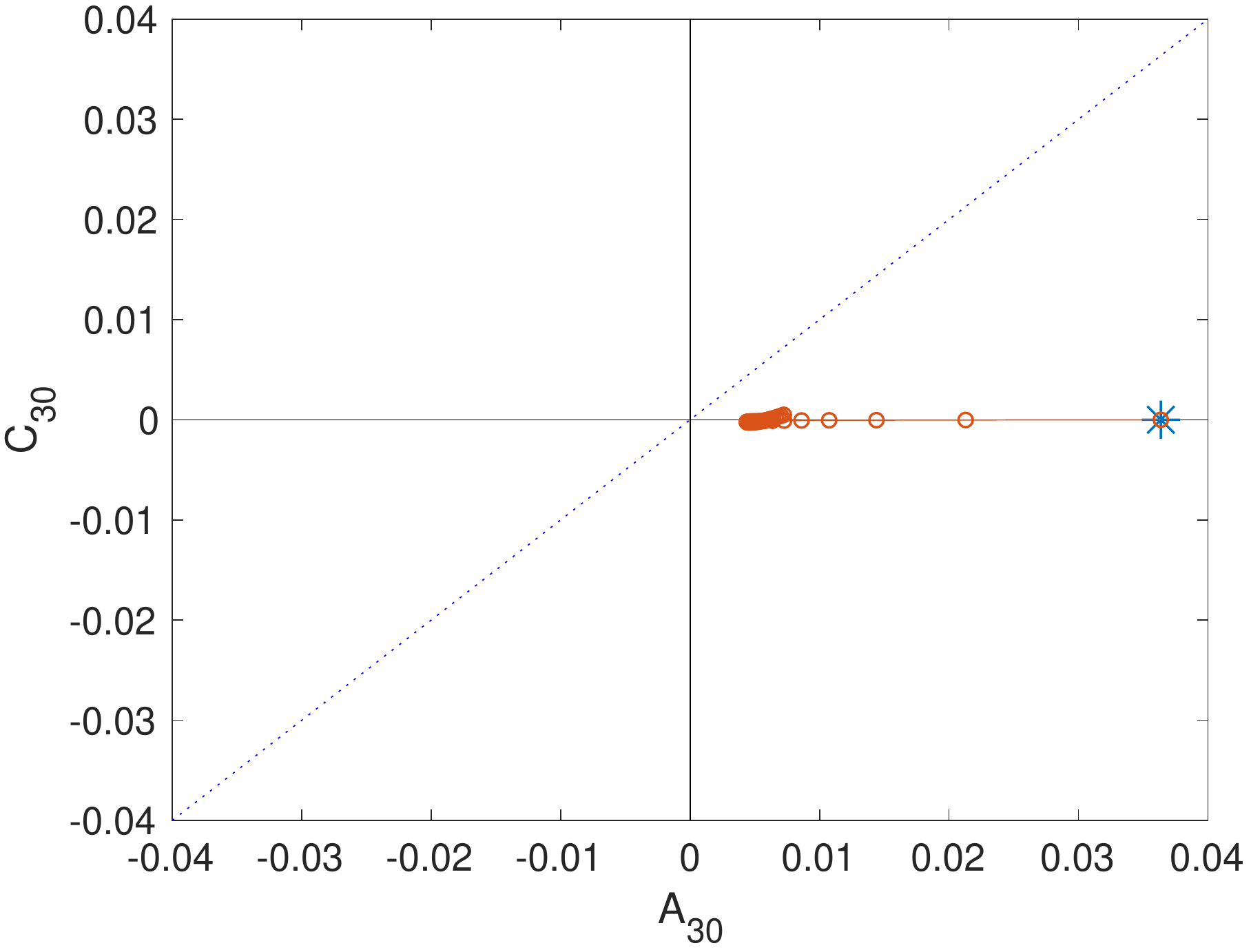}
  \caption{}
  \label{fig:4Layer_outer_AC}
\end{subfigure}
\caption{Plots of the amplitude of interfacial disturbances for four-layer flow in the case that only the inner interface is disturbed initially. Plot (a) shows the amplitude of the disturbance on the inner interface ($A_{30}$) versus the amplitude of the disturbance on the middle interface ($B_{30}$), plot (b) shows the middle interface ($B_{30}$) versus the outer interface ($C_{30}$), and plot (c) shows the inner interface ($A_{30}$) versus the outer interface ($C_{30}$). The parameter values are $Q = 500$, $\mu_i = 0.2$, $\mu_1 = 0.4$, $\mu_2 = 0.6$, $\mu_o = 1$, $T_0 = T_1 = T_2 = 1$, $R_0(0) = 9/11$, $R_1(0) = 10/11$, $R_2(0) = 1$, $n = 30$, $A_{30}(0) = 0.04$, $B_{30}(0) = 0$, $C_{30}(0) = 0$, $0 \leq t \leq 0.01$.}
\label{fig:4Layer_outer}
\end{figure}

\subsection{Time-dependent Injection Rate: Maximum Injection Rate for a Stable Flow}
We now consider flows with a time-dependent injection rate $Q(t)$. In this section we investigate the maximum value of the injection rate that results in a stable flow, which is denoted $Q_M$. This was studied analytically in \S\ref{sec:TimeDepQ}. For two-layer radial flow, the maximum stable injection rate for a given wave number $n$ and radius $R$ of the circular interface is given exactly by the expression \eqref{MaxStableQ_2Layer}. $Q_M$ is found by taking the minimum value over all wave numbers. 

For flows with three or more layers, a lower bound on $Q_M$ is calculated by minimizing the expression \eqref{3-LayerLB} or \eqref{MaxStableQ_NLayer} over all integer values of $n$ and with the initial positions of the interfaces $R_j(0)$. Then the interfacial positions $R_j(\Delta t)$ at the next time step $\Delta t$ are calculated using this injection rate and the process is repeated. To find $Q_M$ exactly for three-layer flow (see Figure \ref{fig:QvsT_Bounds}), an additional calculation is needed. There is no analytical expression for the maximum stable injection rate for given values of $n$, $R_0$, and $R_1$ analogous to equation \eqref{MaxStableQ_2Layer}. Therefore, we use the expression for $\sigma^+(t)$ given in \cite{gin-daripa:hs-rad}. A root finding method is used to find the value of $Q$ such that $\sigma^+(t) = 0$.

The maximum injection rate for which a certain two-layer flow is stable is given by the solid line in Figure \ref{fig:QvsT_Bounds}. As a comparison, the maximum stable injection rate is calculated as discussed above from $\sigma^+(t)$ for a three-layer flow in which the outer interface starts in the same position as the interface for two-layer flow, the viscosity of the inner and outer layers are the same as the two-layer flow, and the intermediate layer has a viscosity which is greater than the inner layer and smaller than the outer layer. This is the dashed line in Figure \ref{fig:QvsT_Bounds}. Note that the three-layer flow is stable for a much larger injection rate due to the fact that the viscosity jumps at the interfaces are smaller.
\begin{figure}
\centering
\includegraphics[width=2.5 in,height=2 in]{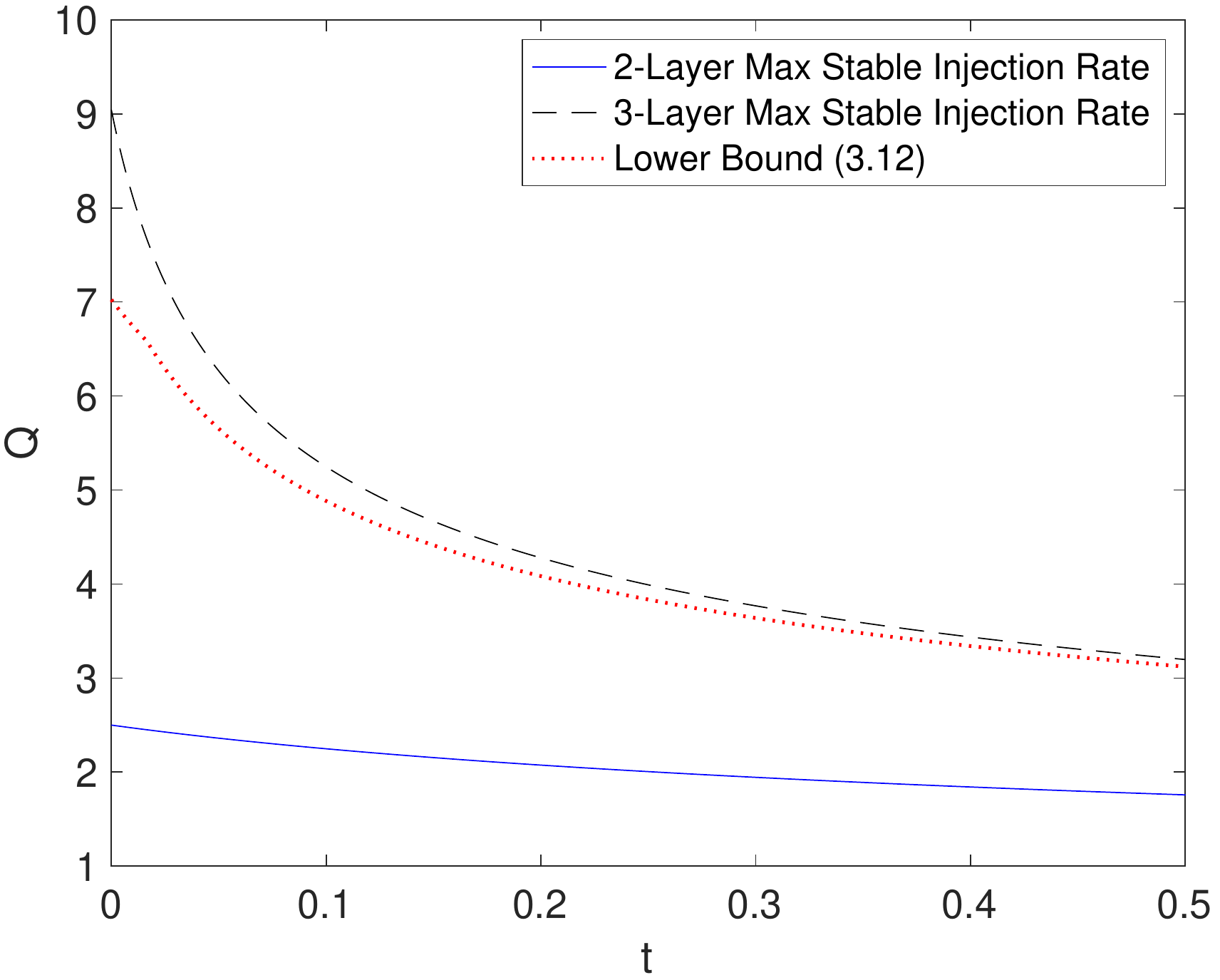}
\caption{Plots of the maximum stable injection rate vs time for two-layer flow (solid line) and three-layer flow (dashed line) as well as the lower bound for stabilization given by equation \eqref{3-LayerLB} (dotted line). The values of the parameters are $\mu_i = 0.2$, $\mu_o = 1$, $T = 1$, $R(0) = 1$ for two-layer flow and 
$\mu_i = 0.2, \mu_1 = 0.6, \mu_o = 1, T_0 = T_1 = 1, R_0(0) = 0.8, R_1(0) = 1$ for three-layer flow.}
\label{fig:QvsT_Bounds}
\end{figure}
Also included in Figure \ref{fig:QvsT_Bounds} is the lower bound on the maximum stable injection rate given in equation \eqref{3-LayerLB}. This is the dotted line in Figure \ref{fig:QvsT_Bounds}. Note that for these particular values of the parameters, this bound, while not strict, allows for a significant increase in the injection rate over two-layer flow.

\begin{figure}
\centering
\includegraphics[width=2.5 in,height=2 in]{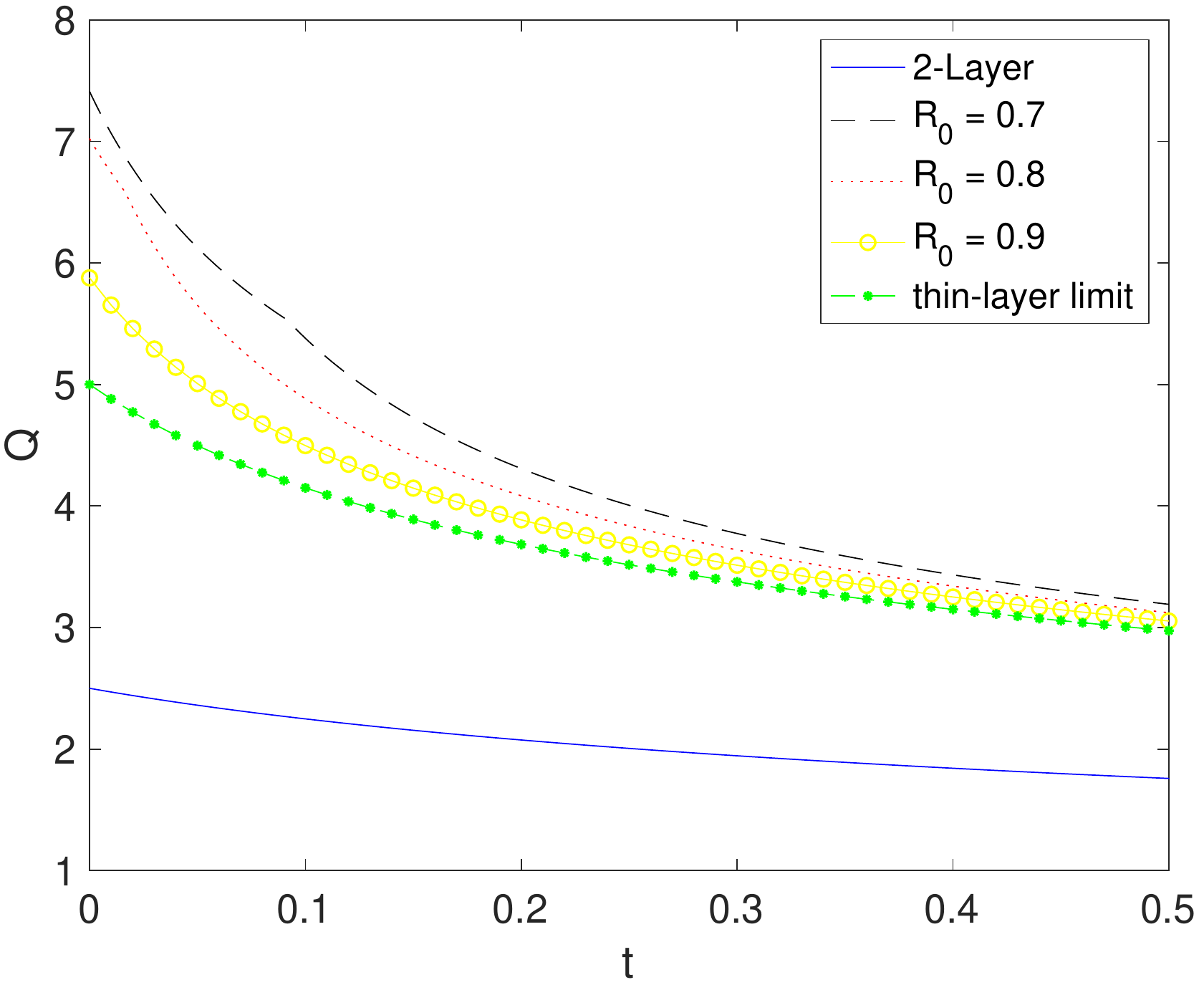}
\caption{Plots of the maximum stable injection rate vs time for two-layer flow (solid line) and three-layer flow (dashed line) as well as the lower bound \eqref{3-LayerLB} for three values of $R_0$. The values of the parameters are $\mu_i = 0.2$, $\mu_o = 1$, $T = 1$, $R(0) = 1$ for two-layer flow and 
$\mu_i = 0.2, \mu_1 = 0.6, \mu_o = 1, T_0 = T_1 = 1, R_0(0) = 0.7,0.8,0.9, R_1(0) = 1$ for three-layer flow.}
\label{fig:QvsT_2vs3Layer}
\end{figure}
In oil recovery applications, it is often expensive to include a more viscous intermediate fluid instead of just using water to displace oil. Therefore, it would be economically advantageous if a minimal amount of fluid could be used in the intermediate layer of a three-layer flow if it still allows the flow to be stabilized at a faster injection rate. 
This behavior is investigated in Figure \ref{fig:QvsT_2vs3Layer}. The solid line is the maximum stable injection rate for two-layer flow using the same parameters as in Figure \ref{fig:QvsT_Bounds}. This flow is compared with three-layer flows for which the inner interface is initially at $R_0(0) = 0.7,0.8,0.9$. Notice that as the middle layer becomes thinner, the maximum stable injection rate decreases. Recall from \S\ref{sec:LimitingCases} that in the limit of an infinitely thin middle layer, the flow will be stable when $Q$ satisfies \eqref{ThinLimit}. The injection rate obtained by taking the minimum value of $Q$ over all $n$ from equation \eqref{ThinLimit} is also shown as ``thin-layer limit'' in Figure~\ref{fig:QvsT_2vs3Layer}. Equation \eqref{ThinLimit} is the same condition as the stable injection rate for two-layer flow but with effective interfacial tension $T_0 + T_1$. Therefore, a three-layer flow with a very thin intermediate layer 
will be stable for faster injection rates than the corresponding two-layer flow as long as the sum of the interfacial tensions in the three-layer flow is greater than the interfacial tension of the two-layer flow. Note that this does \textbf{not} depend on the viscosity of the intermediate fluid. The fact that the middle layer can be very small leads to the conclusion that the use of a thin layer of spacer fluid with desirable properties can be a viable injection strategy.

\begin{figure}
\begin{subfigure}[b]{.5\textwidth}
  \centering
  \includegraphics[width=2.5 in,height=2 in]{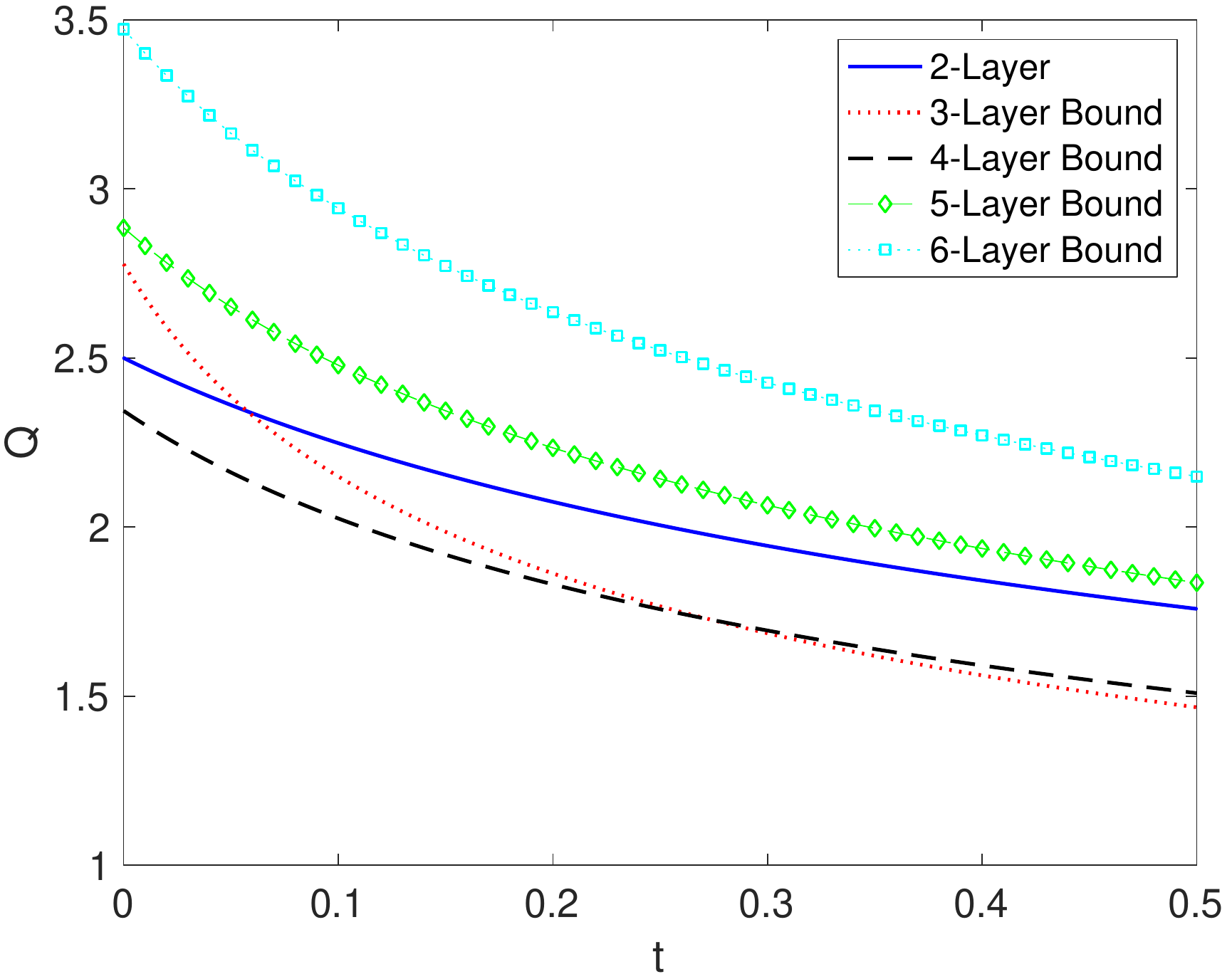}
  \caption{}
  \label{fig:MaxStableQ_NLayerBounds}
\end{subfigure}%
\begin{subfigure}[b]{.5\textwidth}
  \centering
  \includegraphics[width=2.5 in,height=2 in]{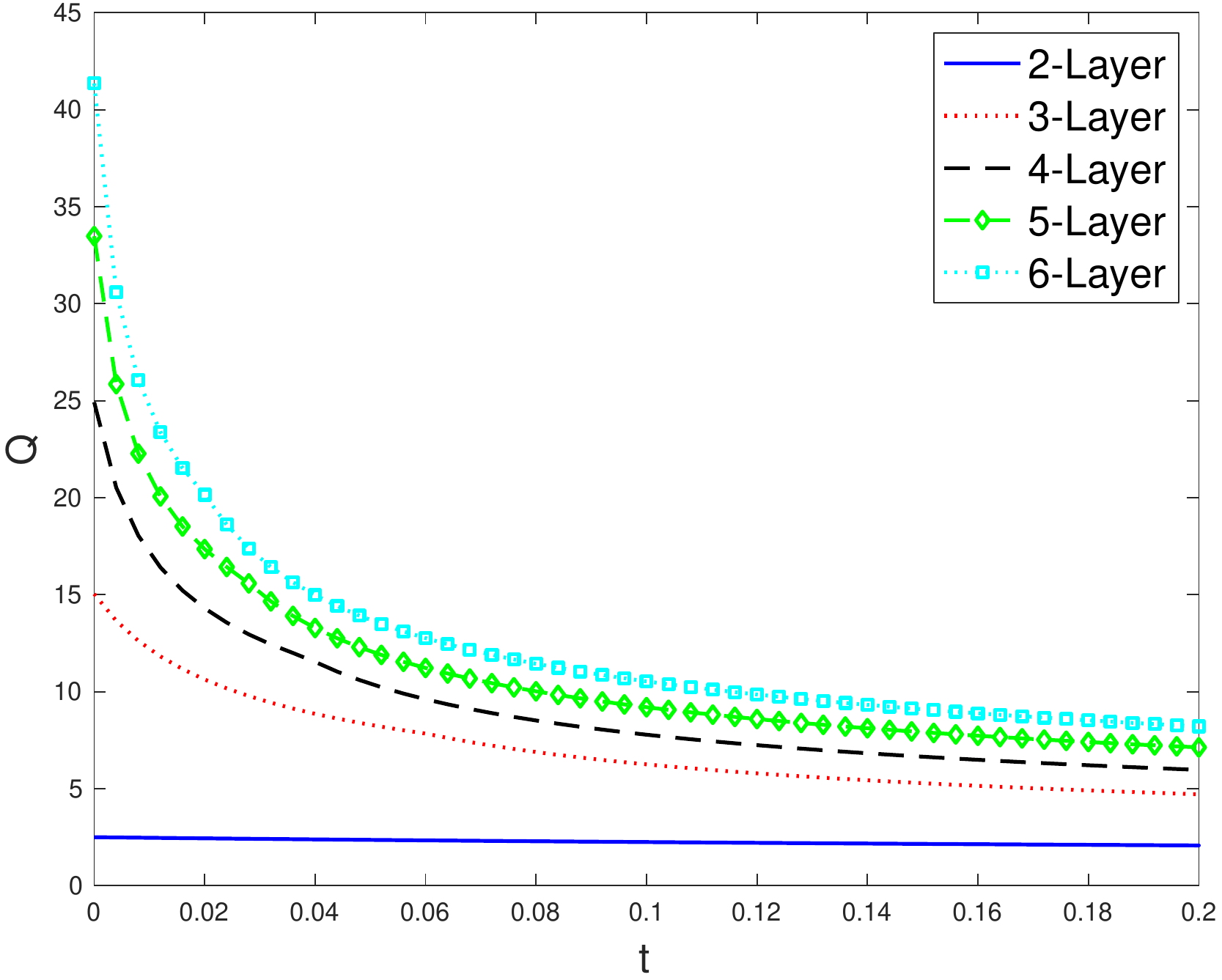}
  \caption{}
  \label{fig:MaxStableQ_NLayer_Shorttime}
\end{subfigure}
\caption{Plots of the maximum stable injection rate vs time for flows with two through six fluid layers. Plot (a) uses the lower bound given by equation \eqref{MaxStableQ_NLayer} for flows with three or more layers while plot (b) shows the actual maximum stable injection rate. The values of the parameters are $\mu_i = 0.2$, $\mu_o = 1$, $T_j = 1$ for all $j$, $R_0(0) = 0.6$, and $R_N(0) = 1$. The interfaces are equally spaced at time $t = 0$ in all cases and the viscous jumps are the same at all interfaces.}
\label{fig:MaxStableQ_NLayer}
\end{figure}

We next compare estimates of the maximum stable injection rates using the lower bound \eqref{MaxStableQ_NLayer} in order to explore how the bound changes as the number of layers increases.
Recall that \eqref{MaxStableQ_NLayer} gives a lower bound on the maximum stable injection rate for a flow with $(N+2)$ layers (or $N$ internal layers). This bound was computed for flows with three, four, five, and six layers and plotted in Figure \ref{fig:MaxStableQ_NLayerBounds}. For comparison, the maximum injection rate which results in a stable flow is plotted for two-layer flow (see equation \eqref{MaxStableQ_2Layer}). For all flows with more than one interface, 
the innermost interface has an initial position of $R_0(0) = 0.6$ and the other interfaces are evenly spaced at time $t = 0$. For all flows, the innermost fluid has a viscosity of $\mu_i = 0.2$, the outermost fluid has a viscosity of $\mu_o = 1$, and the viscosities of all intermediate layers are chosen so that the viscous jump at each interface is the same. All interfaces have the same interfacial tension. In general, the addition of more fluid layers increases the lower bound on the maximum stable injection rate. Intuitively, this is because the jumps at the interfaces are smaller when there are more layers of fluid. The one exception is three-layer flow which has a larger lower bound than four-layer flow for short time. This is due to the fact that for three-layer flow, the lower bound on the maximum stable injection rate only includes the first and last term of equation \eqref{MaxStableQ_NLayer}, which has a different structure than the intermediate terms. For flows with four or more layers and these parameter values, the intermediate terms produce the minima.

In addition to considering approximations to the maximum stable injection rate from lower bounds as discussed above, we also numerically compute the maximum stable injection rate for multi-layer flows. For these computations, a root finding algorithm is used to find the value of $Q$ that results in the maximum eigenvalue of the matrix $\mathbf{M}_N$ (defined in \eqref{M}) being zero. Figure \ref{fig:MaxStableQ_NLayer_Shorttime} shows the maximum stable injection rates for flows with two, three, four, five, and six layers. The parameters used are the same as in Figure \ref{fig:MaxStableQ_NLayerBounds}. A comparison of Figures \ref{fig:MaxStableQ_NLayerBounds} and \ref{fig:MaxStableQ_NLayer_Shorttime} shows that the increase in the maximum stable injection rate from using additional layers of fluid is much greater than what is suggested by the lower bounds. This is especially true for short times. For example, for these values of the parameters the maximum stable injection rate at time $t = 0$ is more than 16 times greater for six-layer flow than it is for two-layer flow. 

\begin{figure}
\centering
\includegraphics[width=2.5 in,height=2 in]{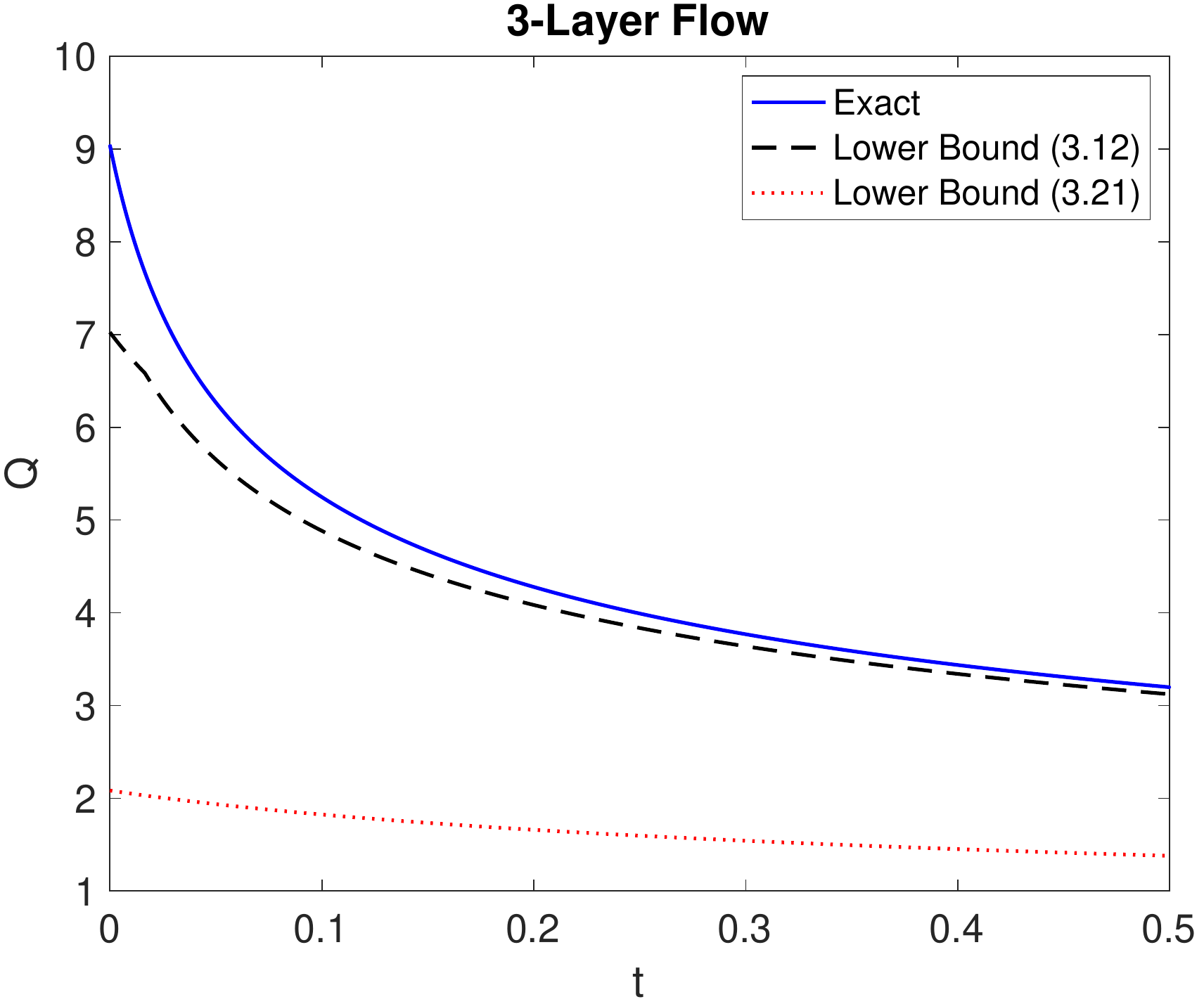}
\caption{Plot of the maximum stable injection rate vs time for three-layer flow (solid line) as well as the lower bounds given by equations  \eqref{3-LayerLB} and \eqref{MaxStableQ_NLayer}. The values of the parameters are $\mu_i = 0.2, \mu_1 = 0.6, \mu_o = 1, T_0 = T_1 = 1, R_0(0) = 0.8, R_1(0) = 1$.}
\label{fig:QvsT_3Layer_Bounds}
\end{figure}

A comparison of Figures \ref{fig:MaxStableQ_NLayerBounds} and \ref{fig:MaxStableQ_NLayer_Shorttime} shows that the lower bounds given by equation \eqref{MaxStableQ_NLayer} are crude. There is the potential to find analytical bounds which are sharper. In particular, the three-layer lower bound \eqref{3-LayerLB} found using Gershgorin's Circle Theorem gives a sharper bound than \eqref{MaxStableQ_NLayer}. One reason that the bound given by equation \eqref{3-LayerLB} is better is because the interfaces remain coupled whereas in the bounds given by  \eqref{MaxStableQ_NLayer} the interfaces have been decoupled. The bound must become cruder in order to decouple the interfaces. Figure \ref{fig:QvsT_3Layer_Bounds} shows a comparison of the two bounds. The solid curve shows the exact value of the maximum stable injection rate for the same three-layer flow considered in Figure \ref{fig:QvsT_Bounds}. The dashed line is the lower bound given by equation \eqref{3-LayerLB} and the dotted line is the lower bound given by equation \eqref{MaxStableQ_NLayer}. The difference between the two bounds is stark.

Recall from Section \ref{sec:Q(t)_2-layer} that Beeson-Jones and Woods \cite{Beeson-Jones/Woods:2015} showed that for two-layer flow (i.e. single interface), the maximum stable injection rate scales like $t^{-1/3}$ for $t >> 1$. This also holds true for flows with multiple interfaces. For flows with 1 through 30 interfaces, we calculated the maximum stable injection rate from $t = 0$ to $t = 100$ and then fit an exponential function of the form $Q(t) = C t^{\alpha}$ for $t \geq 10$. As in Figure \ref{fig:MaxStableQ_NLayer_Shorttime}, the initial position of the innermost interface is $R = 0.6$ and the initial position of the outermost interface is $R = 1$. The interfaces are equally spaced and the viscous jumps at the interfaces are all the same with $\mu_i = 0.2$ and $\mu_o = 1$. The interfacial tension is $1$ for every interface. In all cases, the exponent of the best fit exponential function is $\alpha = -1/3$, which matches with the analytically obtained scaling law for the single interface case. However, the constant $C$ increases with the number of interfaces. The values of $C$ are plotted versus the number of interfaces on a log-log scale in Figure \ref{fig:MaxStableQ_NLayer_C}. The line of best fit through the points is shown in the figure and has slope $2/3$. Therefore, if $N_I$ is the number of interfaces, then $C \propto N_I^{2/3}$. Thus $Q(t) \propto N_I^{2/3} t^{-1/3}$ for $t \geq 10$.
\begin{figure}
\centering
\includegraphics[width=2.5 in,height=2 in]{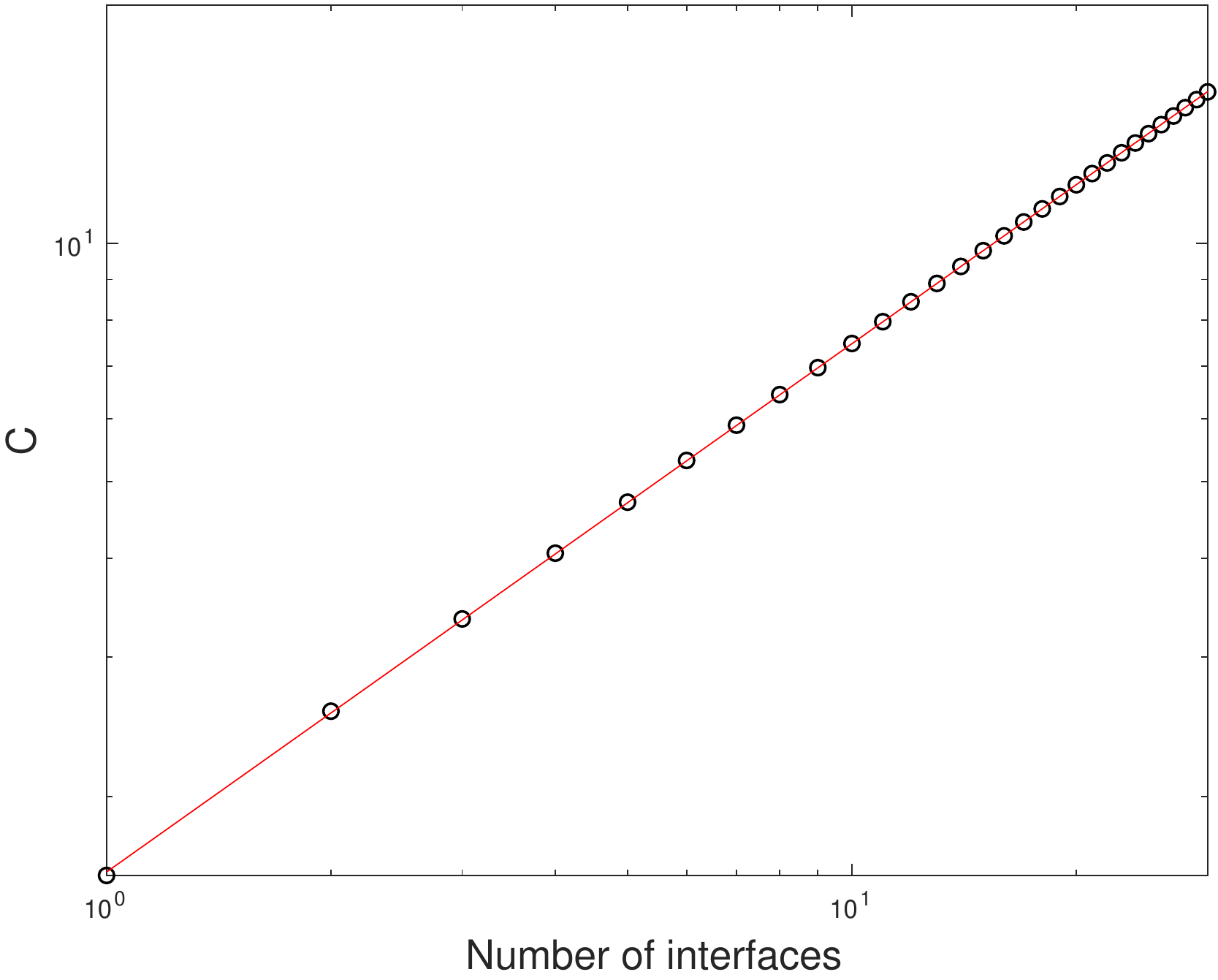}
\caption{A plot of $C$ versus the number of interfaces where the long-time behavior of the maximum stable injection rate is of the form $Q(t) = C t^{-1/3}$. The values of the parameters are $\mu_i = 0.2$, $\mu_o = 1$, $T_j = 1$ for all $j$, $R_0(0) = 0.6$, and $R_N(0) = 1$. The interfaces are equally spaced at time $t = 0$ in all cases and the viscous jumps are the same at all interfaces.}
\label{fig:MaxStableQ_NLayer_C}
\end{figure}

This relationship between $C$ and $N_I$ can be understood in the following way. First, notice that the maximum stable injection rate for flow with a single interface is independent of the interfacial tension $T$ (see equation~\eqref{MaxStableQ_2Layer}). However, the interfacial tension is part of both the characteristic injection rate and the characteristic time scale. The dimensionless injection rate $Q^*$ is proportional to $1/T$ and the dimensionless time $t^*$ is proportional to $T$. Therefore, since $Q^*(t^*) \propto (t^*)^{-1/3}$, 
\begin{equation*}
 \frac{Q(t)}{T} \propto (Tt)^{-1/3} \implies Q(t) \propto T^{2/3} t^{-1/3}.
\end{equation*}
Recall from Section \ref{sec:LimitingCases} that when there are two interfaces, they will eventually be very close together. In that limit, the maximum stable injection rate reduces to a term that is identical to a single interface maximum stable injection rate but with interfacial tension equal to the sum of the interfacial tensions of the two interfaces. Therefore, if both interfaces have the same interfacial tension, the maximum stable injection rate with two interfaces will be greater than the comparable single interface flow by a factor of $2^{2/3}$. For $N_I$ interfaces, the long-time behavior is the same as the single interface case where the single interface has interfacial tension equal to the sum of the $N_I$ interfacial tensions. Therefore, if all of the interfaces have the same interfacial tension we would expect that $Q(t) \propto N_I^{2/3}$ which agrees very well with the results of Figure \ref{fig:MaxStableQ_NLayer_C}.

In summary, the injection rate for a stable flow increases at a rate proportional to the number of interfaces to the two-thirds power at large time $t >> 1$. However, at earlier times the number of interfaces can increase the maximum stable injection rate by a much greater amount.

\subsection{Time-dependent Injection Rate: Stable Manifold}
Finally, we numerically investigate the analytical findings of \S\ref{Qt_for_F0} where it was analytically proven that for an appropriate time-dependent injection rate, if only the inner interface is disturbed initially then the outer circular interface will remain circular for all time and the disturbance on the inner interface will decay. A phase plane trajectory of such an initial data consisting of a disturbance of amplitude $10^{-2}$ on the inner circular interface with no disturbance on the outer circular interface is shown in Figure~\ref{fig:F0_0} and is labeled as ``stable''. The axes of the phase plane are $A_{20}$ and $B_{20}$ corresponding to the amplitudes of disturbances on the inner and the outer circular interfaces respectively. Notice that the trajectory is along the horizontal axis towards the origin as expected. However, when the outer circular interface is also perturbed by a disturbance of infinitesimal amplitude $10^{-6}$, $10^{4}$ times smaller than the one on the inner circular interface, and keeping the time dependent injection rate the same, the dynamics are markedly different which is shown by the trajectory marked ``unstable'' in this figure. Notice that for short time only the disturbance on the inner interface decays before the disturbances of both the inner and outer interfaces begin to grow. Therefore, a very small difference in the initial data can lead to wildly different behavior at later times.

Figures~\ref{fig:F0_0_Interfaces_t0} and \ref{fig:F0_0_Interfaces_t300} respectively show the interface positions at the beginning and end times for the ``stable'' and ``unstable'' phase trajectories shown in Figure~\ref{fig:F0_0}.
The initial set-up of the interfaces in stable trajectory is so close to that in the unstable trajectory that the difference can not easily be discerned in the figures (see Figure \ref{fig:F0_0_Interfaces_t0}). However, at later times there is a clear difference (see Figure \ref{fig:F0_0_Interfaces_t300}). At time $t = 0.001$, the outer interface that corresponds to the ``stable" trajectory is circular. However, there is clearly a disturbance in the case of the ``unstable'' trajectory. 

In summary, an appropriate injection rate can result in a stable configuration which consists of all perfectly circular interfaces except for one which is perturbed with a monochromatic wave of infinitesimal amplitude. The prescribed injection rate results in the perturbation decaying, even if the configuration (i.e. the set-up consisting of all but one circular interfaces and one perturbed circular interface) itself is unstable to disturbances on any one or more of the circular interfaces. In practice, it may be difficult to realize such a flow.

\begin{figure}
\centering
\includegraphics[width=2.5 in,height=2 in]{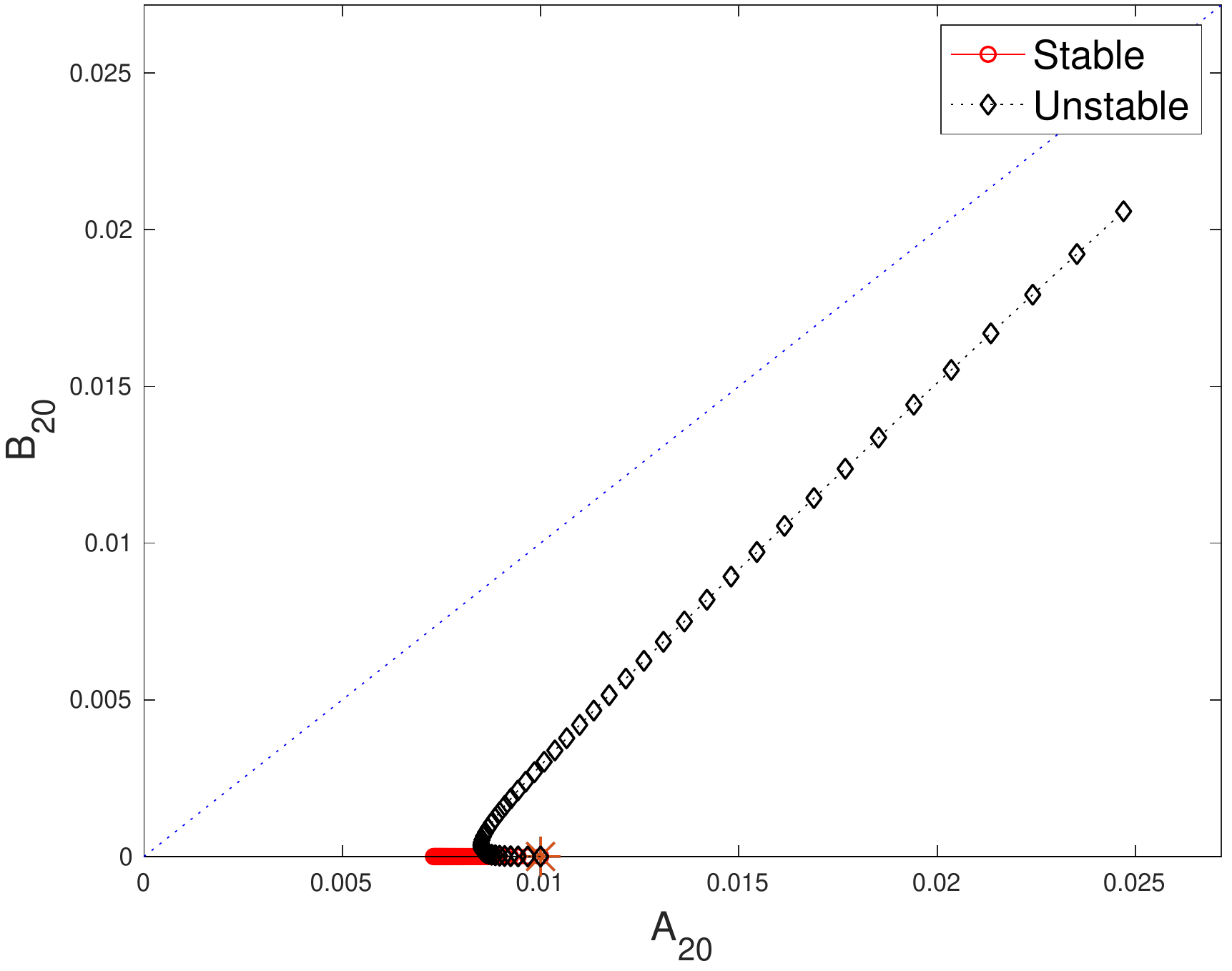}
\caption{Plots of the amplitude of the disturbance of the inner interface ($A_{20}(t)$) versus the amplitude of the disturbance of the outer interface ($B_{20}(t)$) when the injection rate is given by the formula \eqref{Q_F0}, $Q(t) = T_0(n^2-1)/(6R_0(t)(\mu_1-\mu_i))$. The curve labeled ``stable'' is the trajectory of initial data $(A_{20}(0), B_{20}(0))=(10^{-2},0)$ which corresponds to a perturbed inner interface and circular outer interface. The curve labeled ``unstable'' is the trajectory of initial data $(A_{20}(0), B_{20}(0))=(10^{-2},10^{-6})$ which corresponds to a perturbed inner interface and a slightly perturbed outer interface.
The parameter values are $\mu_i = 0.2, \mu_1 = 0.3, \mu_o = 1, T_0 = 15, T_1 = 1, R_0(0) = 14/15, R_1(0) = 1, n = 20$, $0 \leq t \leq 10^{-3}$.}
\label{fig:F0_0}
\end{figure}

\begin{figure}
\begin{subfigure}[b]{.5\textwidth}
  \centering
  \includegraphics[width=2.5 in,height=2 in]{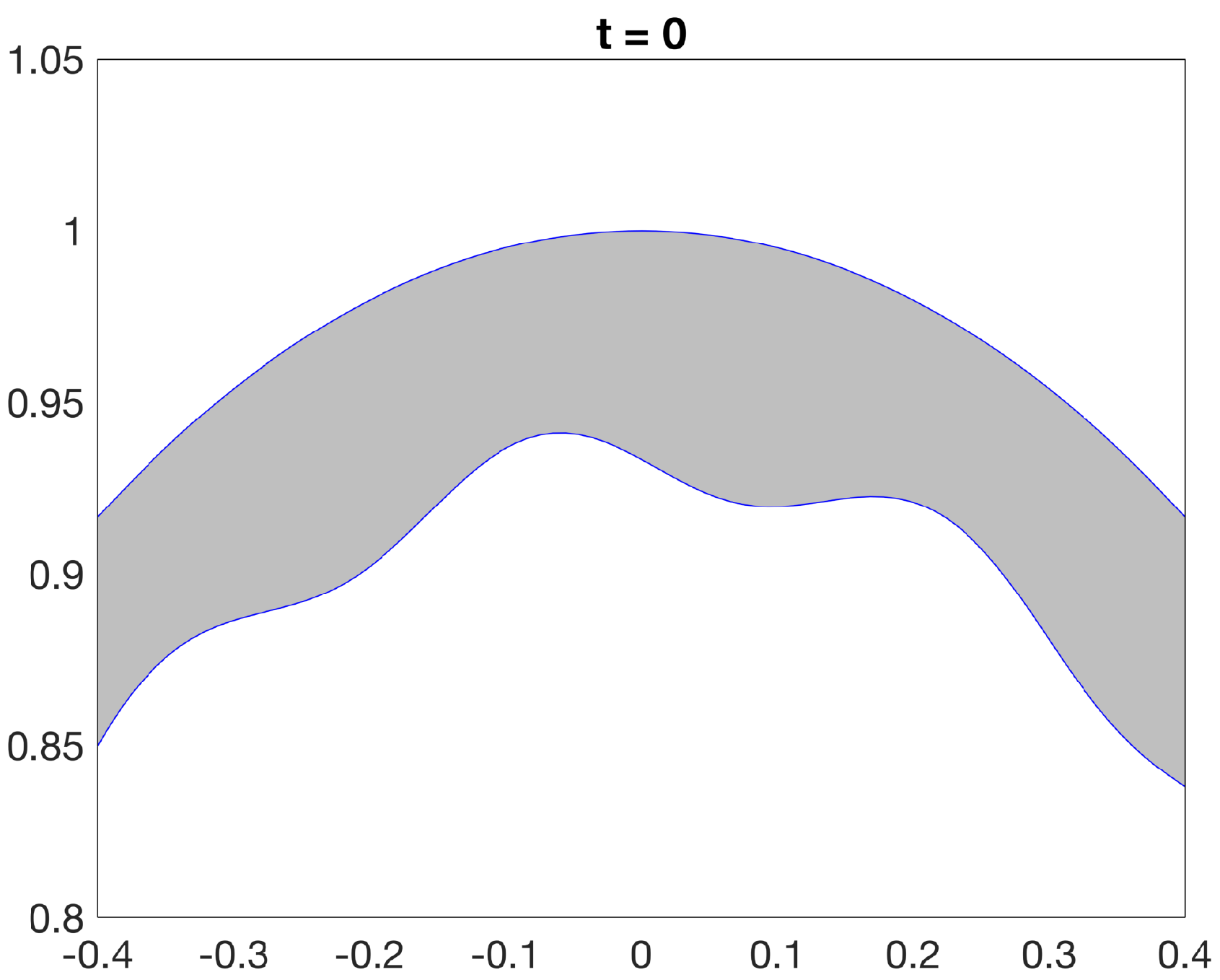}
\end{subfigure}%
\begin{subfigure}[b]{.5\textwidth}
  \centering
  \includegraphics[width=2.5 in,height=2 in]{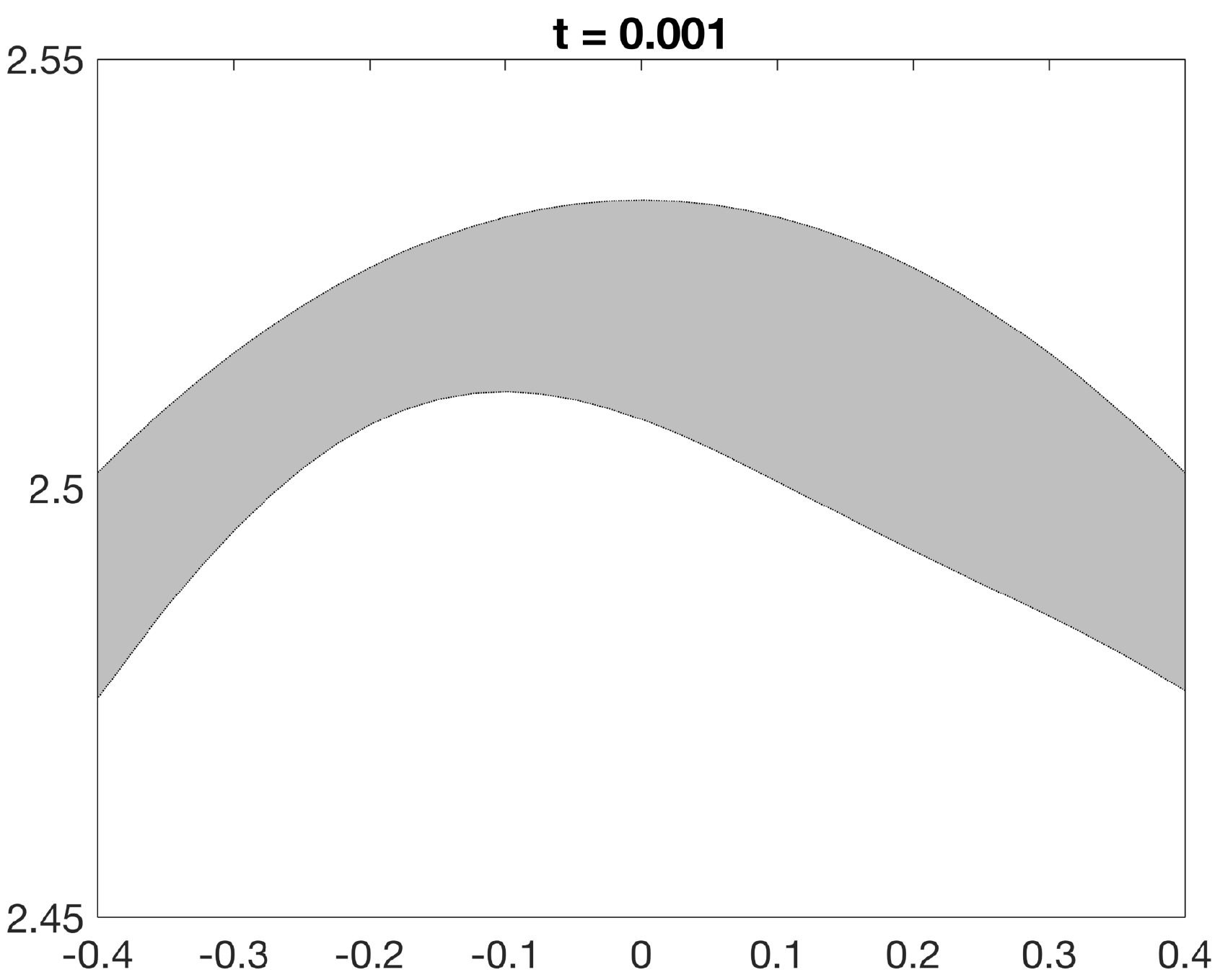}
\end{subfigure}
\begin{subfigure}[b]{.5\textwidth}
  \centering
  \includegraphics[width=2.5 in,height=2 in]{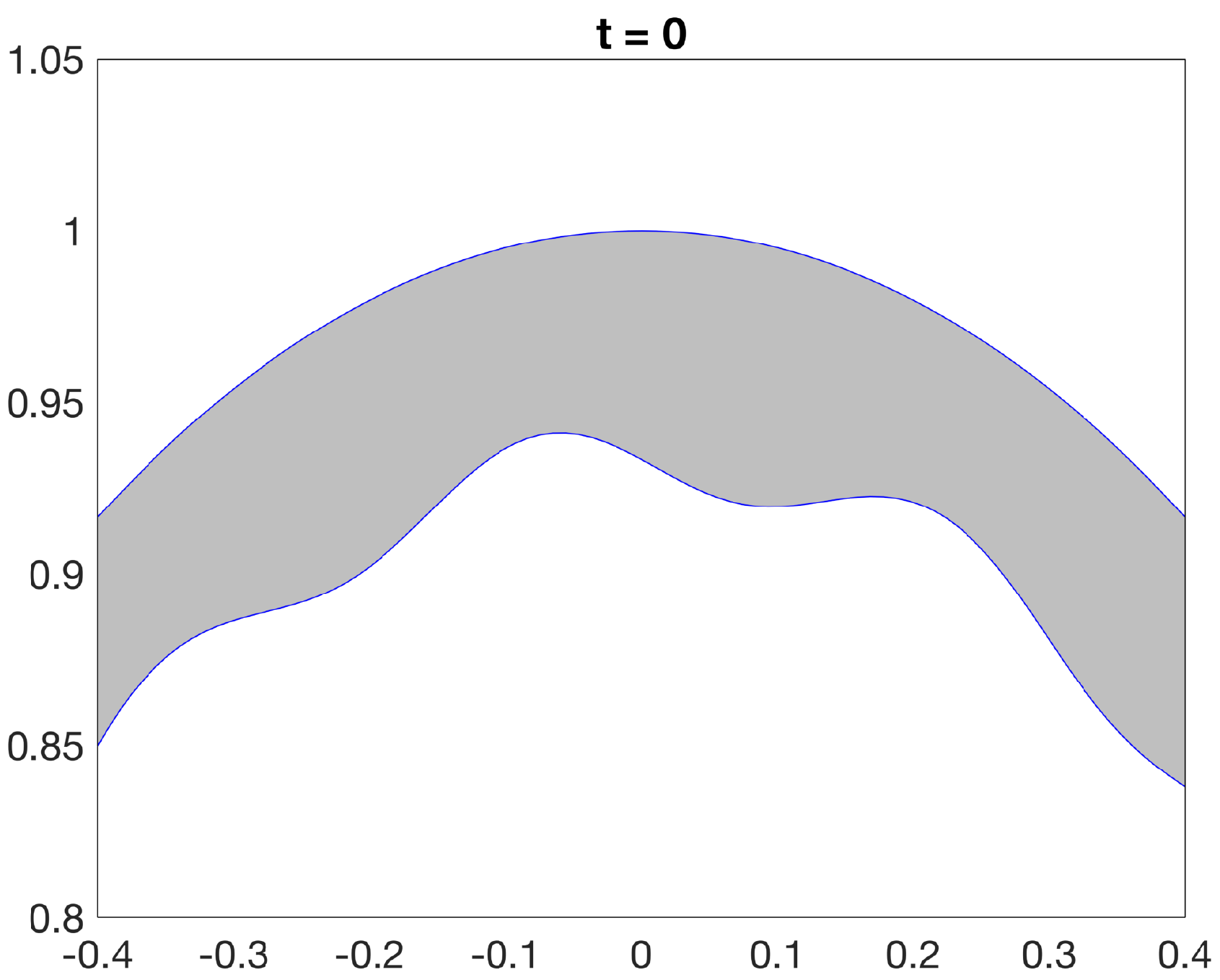}
  \caption{}
  \label{fig:F0_0_Interfaces_t0}
\end{subfigure}%
\begin{subfigure}[b]{.5\textwidth}
  \centering
  \includegraphics[width=2.5 in,height=2 in]{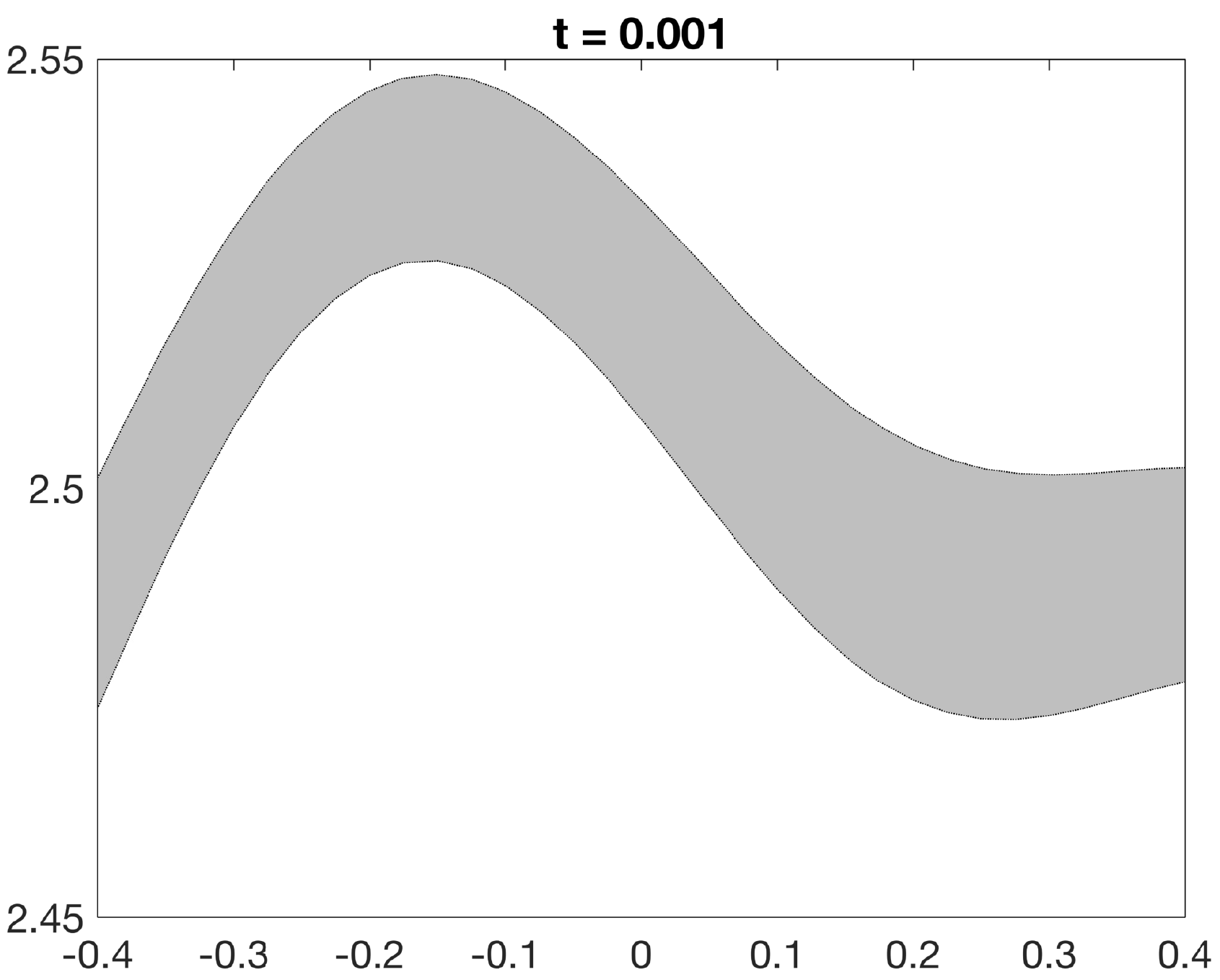}
  \caption{}
  \label{fig:F0_0_Interfaces_t300}
\end{subfigure}
\caption{Plots of the interfaces when the injection rate is given by the formula \eqref{Q_F0} (see Figure \ref{fig:F0_0}) at times (a) $t = 0$ and (b) $t = 0.001$. The top row shows the interfaces corresponding to the ''stable'' trajectory in Figure \ref{fig:F0_0}, and the bottom row shows the interfaces corresponding to the ''unstable'' trajectory.}
\label{fig:F0_0_Interfaces_zoom}
\end{figure}

\section{Conclusion}\label{sec:Conclusion}

In this paper, the linear stability of multi-layer immiscible Hele-Shaw and porous media flows has been investigated. The system of ODEs that governs the evolution of linearized interfacial disturbances has been derived for flows with an arbitrary number of fluid layers. Several key results about time-dependent injection rates are obtained which can be used to stabilize the flow. In applications like oil recovery, it can be advantageous to inject fluid as quickly as possible while maintaining a stable flow.
For two-layer flows, the maximum injection rate for which a disturbance with a particular wave number is stable is given in \cite{Beeson-Jones/Woods:2015}. A similar condition is found in the present work for three-layer flows by using Gershgorin's circle theorem.  
By using upper bounds derived in \cite{gin-daripa:hs-rad} a sufficient condition on the injection rate to ensure stability is found for a flow with an arbitrary number of fluid layers. 

The behavior of the interfaces and the growth rates of disturbances are investigated numerically and the following results are found: 
(a) The interfaces are coupled so that a disturbance on one interface is transferred to the other interface(s); (b) The disturbances on the interfaces can be in phase or out of phase and there can be a transition from one state to the other; (c) The addition of extra fluid layers can have a large impact on the dynamics of the interfacial motion; (d) Flows with more fluid layers can be stable with faster injection rates than comparable flows with fewer fluid layers, and the maximum stable injection rate is proportional to the number of interfaces raised to the two-thirds power; and (e) The use of an appropriate injection rate can result in a stable configuration which consists of all perfectly circular interfaces except for one which is perturbed with a monochromatic wave of infinitesimal amplitude. The prescribed injection rate results in the perturbation decaying, even if the configuration (i.e. the set-up consisting of all but one circular interfaces and one perturbed circular interface) itself is unstable to disturbances on any one or more of the circular interfaces. In practice, it may be difficult to realize such a flow. These results are independent of viscosities of the fluids and the values of interfacial tension as long as the viscosity jump is positive in the direction of flow at each interface.

\section*{\bf Acknowledgments:}
This work has been supported in part by the U.S. National Science Foundation grant DMS-1522782.
\bigskip

\pagebreak

\clearpage
\bibliographystyle{siam}
\bibliography{references-HS-EOR}

\end{document}